\documentclass[a4paper,11pt]{article}
\pdfoutput=1 % if your are submitting a pdflatex (i.e. if you have
\usepackage{jcappub} % for details on the use of the package, please
\usepackage{xcolor}
\usepackage[T1]{fontenc} % if needed

\title{\boldmath Inflation is always semi-classical: \\ Diffusion domination overproduces Primordial Black Holes}

 \author{G. Rigopoulos}
 \author{and A. Wilkins} 
 \affiliation{School of Maths, Stats \& Physics, \\Newcastle University,\\Newcastle, UK}

\emailAdd{gerasimos.rigopoulos@newcastle.ac.uk}
\emailAdd{a.wilkins3@newcastle.ac.uk}
\abstract{We use the Hamilton-Jacobi (H-J) formulation of stochastic inflation to describe the evolution of the inflaton during a period of Ultra-Slow Roll (USR), taking into account the field's velocity and its gravitational backreaction. We demonstrate how this formalism allows one to modify existing slow-roll (SR) formulae to be fully valid outside of the SR regime. We then compute the mass fraction, $\beta$, of Primordial Black Holes (PBHs) formed by a plateau in the inflationary potential. By fully accounting for the inflaton velocity as it enters the plateau, we find that PBHs are generically overproduced before the inflaton's velocity reaches zero, ruling out a period of free diffusion or even stochastic noise domination on the inflaton dynamics. We also examine a local inflection point and similarly conclude that PBHs are overproduced before entering a quantum diffusion dominated regime. We therefore surmise that the evolution of the inflaton is always predominantly classical with diffusion effects always subdominant. Both the plateau and the inflection point are characterized by a very sharp transition between the under- and over-production regimes. This can be seen either as severe fine-tunning on the inflationary production of PBHs, or as a very strong link between the fraction $\beta$ and the shape of the potential and the plateau's extent.  }

\begin{document}
\maketitle
\flushbottom

\section{Introduction}
\label{sec:intro}
Primordial Black Holes (PBHs) were first theorised in the $60$s and $70$s \cite{1967SvA....10..602Z, Hawking1971} and it was soon realised that they could be a Dark Matter (DM) candidate \cite{Hawking1971, CHAPLINE1975}. Interest in PBHs has been renewed in the wake of the LIGO-VIRGO detection of the merger of intermediate mass black holes \cite{Abbott2016} which could be primordial rather than astrophysical in origin \cite{Bird2016, Sasaki2016, Clesse2017}. There are numerous constraints on the abundance of PBHs from many different effects -- for a comprehensive review of these see e.g. \cite{Carr2020} or for a shorter pedagogical overview see e.g. \cite{Green2020}. Even if PBHs are not all of DM, their abundance (be it small or large) can act as an invaluable probe of the inflationary potential outside of the narrow CMB window.  

If one wants to generate an appreciable number of PBHs from single-field inflation then one generally needs to go beyond the Slow-Roll (SR) regime into a so-called period of Ultra-Slow Roll (USR) inflation \cite{Tsamis2004, Kinney2005, Namjoo2013, Martin2013, Dimopoulos2017, Salvio2018, Pattison2018}. A period of USR is characterised by a negligible gradient in the potential $V_{,\phi} \sim 0$ or, equivalently, the second SR parameter $\epsilon_2 \sim 6$\footnote{Different conventions will define this slightly differently, for instance $\epsilon_2 \sim \pm 6$ or $\eta \sim \pm 3$}. While there has been a lot of work done on generating PBHs from this period of inflation \cite{Germani2017, Pattison2017, Biagetti2018, Ezquiaga2018, Firouzjahi2019, Passaglia2019, Ezquiaga2020}, until very recently most work focused on the large velocity -- e.g. \cite{Firouzjahi2019} -- or negligible velocity/diffusion dominated regime -- e.g \cite{Pattison2017, Ezquiaga2020}. While there have been strong efforts to describe both limits in the same framework \cite{Pattison2021}, there isn't good control over the transition period between the two regimes. Direct numerical simulation of the stochastic equations of motion is usually prohibitively expensive to get accurate values for the PBH mass fraction -- see \cite{Figueroa2020} for a treatment of this problem. In this paper we will follow the Hamilton-Jacobi (H-J) formalism as originally described by Salopek and Bond \cite{Salopek1990} to describe the evolution of the inflaton. One key advantage of this approach is that the dynamics are reduced to first order without making any assumptions about being in a SR or USR regime - the approximation only involves dropping higher order spatial gradient terms (i.e. focuses on long wavelengths) and includes the field's velocity fully. This will enable us to smoothly describe the transition between SR and USR regimes. Stochastic Inflation \cite{Starobinsky1994} has enjoyed much success as the leading\footnote{see e.g. \cite{Celoria2021} for an alternative description} framework to describe the evolution of non-linear perturbations and their backreaction on the dynamics of the inflaton. One of the reasons it is important to go beyond standard, linear cosmological perturbation theory is to describe the formation of black holes formed in the early universe by large density perturbations. The validity of the stochastic approach to USR has been questioned, with \cite{Cruces2019} and \cite{Pattison2019} weighing in on opposing sides of the issue. We sidestep this issue by merging the H-J description with stochastic inflation, as shown in \cite{Prokopec2019}, in order to have a full, non-linear, description of inflationary dynamics which can include large quantum-stochastic backreaction.

The outline of the paper is as follows. In section \ref{sec: PBHs from USR} we describe how stochastic inflation gives us a framework to compute the abundance of PBHs when they are formed, quantified by the mass fraction, $\beta$. We review the stochastic H-J formalism of \cite{Prokopec2019} and extend its results to plateaus of finite width. We conclude by discussing the stochastic $\delta N$ formalism and noting that standard SR formulae can be adapted to be valid outside the SR regime. In section \ref{sec: plateau} we use heat kernel techniques to compute the mass fraction on a plateau over all possible initial conditions. We find that the number of PBHs produced will violate observational and theoretical constraints before the classical velocity of the inflaton reaches zero. This therefore \textit{forbids a period of free diffusion} where the main or only influence on the inflaton dynamics is the stochastic noise term. In section \ref{sec: Inflection} we modify the mass fraction formulae from \cite{Pattison2017} so that they are valid outside of SR to investigate a local inflection point in the potential. In particular we expand around the classical limit and again find that PBHs will be overproduced before entering a quantum diffusion dominated regime. This suggests that diffusion effects are always subdominant on the inflaton's trajectory. We summarise our results in section \ref{sec: Conclusion}. In appendix \ref{sec: More Venin formulae} we adopt some more standard SR formulae in terms of the H-J framework so as to be valid outside of SR. In appendix \ref{sec: SR + USR} we describe how one can combine the probability density functions of a SR and USR distribution to obtain the appropriate mass fraction. In appendix \ref{sec: PDF for scenario B} we derive the probability density function for a period of free diffusion preceded by a H-J trajectory using heat kernel techniques.
\section{\label{sec: PBHs from USR}Seeding PBHs from a period of USR inflation}
The PBH mass fraction, $\beta$, can be computed from the probability distribution function (PDF) of the coarse-grained scalar curvature perturbation $\zeta_{cg}$:
\begin{eqnarray}
\beta (M) &=& 2 \int_{\zeta_{c}}^{\infty} P(\zeta_{cg})~\mathrm{d}\zeta_{cg} \label{eq:massfracdef} \\
\zeta_{cg}(\textbf{x}) &\equiv & (2\pi)^{-3/2} \int_{k > aH_{form}}\mathrm{d}\textbf{k}\zeta_{\textbf{k}}e^{i\textbf{k}\cdot\textbf{x}} \label{eq:zetacg defn}
\end{eqnarray}
We can clearly see that the mass fraction $\beta$ represents the area under the curve\footnote{Multiplied by a factor of 2 to account for the under-counting in Press-Schechter theory  \cite{Press1974}.} of the PDF above some critical value, $\zeta_c$. The precise value of $\zeta_c$ is not known \textit{a priori} but depends on both the equation of state at horizon re-entry and the shape of perturbation itself \cite{Musco2020}. However it is known to be very close to 1 and despite $\beta$ generically being exponentially sensitive to $\zeta_c$ we will see this does not change the qualitative nature of our results. 

The use of the curvature perturbation, $\zeta$, to compute the PBH mass fraction is heavily criticised in the literature \cite{Musco2019, Young2019, Germani2020, Young2020}. It is clear that to get the most accurate result one should instead replace $\zeta$ in (\ref{eq:massfracdef}) with the density contrast $\delta$ which is related to $\zeta$ in a highly non-linear way. However, such non-linear effects are expected to only reduce $\beta$ at most by a factor of a few, eg $\sim 2$ according to \cite{Young2019}. This is only true if the power spectrum is very peaked, as it will be for us, more generic shapes of the power spectrum are analysed in \cite{Germani2020}. As we will see, such theoretical uncertainties will not alter our conclusions and we therefore neglect them here.\footnote{While preparing this manuscript Biagetti \textit{et. al} \cite{Biagetti2021} demonstrated an explicit method for converting between the pdf for $\zeta$ and pdf for $\delta$. We leave the application of this treatment to our pdfs for future work.}  

Many constraints have been placed on the abundance of PBHs of mass, $M_{PBH}$, in the range $(10^{9}-10^{50})$g -- see \cite{Green2020} for a recent review. These limit the upper bound of $\beta$, the mass fraction of PBHs, to $10^{-24}$ – $10^{-17}$ for $10^9$g$ < M_{PBH} < 10^{16}$g and $10^{-11}$ – $10^{-5}$ for $10^{16}$g$ < M_{PBH} < 10^{50}$g. More recently \cite{Papanikolaou2020}, an upper bound on $\beta$, in the range of $10^{-4}$ – $10^{-2}$ has been proposed for $10$g$ < M_{PBH} < 10^9$g. It is these constraints that we will use throughout this work to restrict the period of USR.

While PBHs can be formed from bubble collisions \cite{Crawford1982, Hawking1982, Kodama1982}, cosmic strings \cite{Hawking1989, Polnarev1991} or the collapse of domain walls \cite{Rubin2000, Rubin2001} to name a few, we will focus here on curvature perturbations generated from a period of inflation. We will outline how inflation does this in the rest of this section.

%\subsection{Homogeneous Inflation}
%The most general metric that obeys homogeneity and isotropy is the well known FLRW metric:
%\begin{eqnarray}
%\mathrm{d}s^2 = -\mathrm{d}t^2 + a(t)^2\left[ \mathrm{d}r^2 + r^2 \mathrm{d}\Omega^2 \right] \label{eq: FLRW metric}
%\end{eqnarray}
%where $a(t)$ is known as the scale factor, $t$ is cosmic time and we have assumed a flat universe. For a universe dominated by an inflaton field $\varphi$ (which in turn is dominated by its potential energy over its kinetic energy) the equation of motion for $\varphi$ is given by the Klein-Gordon equation:
%\begin{eqnarray}
%\ddot{\varphi} + 3H\dot{\varphi} + V_{,\varphi} = 0 \label{eq:K-G}
%\end{eqnarray}
%where we have introduced the Hubble expansion rate $H \equiv \dot{a}/a$. This equation is supplemented by the GR energy constraint also known in this context as the Friedmann equation:
%\begin{eqnarray}
%H^2 = \dfrac{1}{3}\left( \dfrac{\Pi^2}{2} + V(\varphi)\right) \label{eq:Friedmann}
%\end{eqnarray}
%For given initial conditions on the field value $\varphi$ and its velocity $\Pi$ equations (\ref{eq:K-G}) and (\ref{eq:Friedmann}) provide a complete description of the homogeneous evolution of the inflaton field for a given potential $V(\phi)$. 

\subsection{Classical Inhomogeneous Inflation}
It can be shown \cite{Salopek1990} that during inflation the long wavelength metric can be written as: 
\begin{eqnarray}
\mathrm{d}s^2 = -N^2(t,x^i)\mathrm{d}t^2 + e^{2\alpha(t,\textbf{x})}h_{ij}(\textbf{x})\mathrm{d}x^i\mathrm{d}x^j \label{eq:longwavemetric}
\end{eqnarray}
The shift vector, $N_i$, has been set to $0$ but coordinate freedom remains in the choice of the lapse function $N$. 
The local expansion rate is defined as:
\begin{eqnarray}
H(t,\textbf{x}) \equiv \dfrac{1}{N}\dfrac{\partial \alpha}{\partial t} \label{eq: inhom H expan rate}
\end{eqnarray}
while the dynamics of $h_{ij}(\textbf{x})$, describing volume-preserving deformations of the spatial geometry, can be ignored as a first approximation. By keeping the leading order in spatial gradients we can obtain a dynamical equation for the long wavelength modes of the inflaton field $\phi$:
\begin{eqnarray}
\Pi = \dfrac{1}{N}\dfrac{\partial \phi}{\partial t} \label{eq: mom defn} \\
\dfrac{1}{N}\dfrac{\partial \Pi}{\partial t} + 3H \Pi + \dfrac{\mathrm{d}V}{\mathrm{d}\phi} = 0 \label{eq: inhom K-G}
\end{eqnarray} 
and we also obtain the energy constraint equation:
\begin{eqnarray}
H^2 = \dfrac{1}{3 M_{\rm p}^2}\left( \dfrac{\Pi^2}{2} + V(\phi) \right) \label{eq: inhom energy constraint}
\end{eqnarray}
It important to realise that although equations (\ref{eq: inhom K-G}) and (\ref{eq: inhom energy constraint}) look identical to the homogeneous versions, %(\ref{eq:K-G}) 
%and 
%(\ref{eq:Friedmann}) 
they are valid at each spatial point with \textit{a priori} different initial conditions. Equations (\ref{eq: inhom K-G}) and (\ref{eq: inhom energy constraint}) represent the \textit{separate universe evolution} as each spatial point independently follows its own homogeneous cosmology evolution. What has yet to be taken into account however is the GR momentum constraint which must also be obeyed and we will see this restricts the separate universe picture.   
\subsubsection{The momentum constraint}
At leading order in spatial gradients, the GR momentum constraint tells us that:
\begin{eqnarray}
\partial_{i}H = -\dfrac{1}{2M_{\rm p}^2}\Pi \partial_{i}\phi \label{eq: GR mom constraint}
\end{eqnarray}
Taking this additional constraint into account, one can show \cite{Salopek1990} that both $H$ and $\Pi$ are solely functions of $\phi$ with no explicit time dependence and are related through:
\begin{eqnarray}
\Pi(\phi) = -2 M_{\rm p}^2 \,\dfrac{\mathrm{d}H(\phi)}{\mathrm{d}\phi} \label{eq: Pi and H relation}
\end{eqnarray}
If (\ref{eq: Pi and H relation}) is inserted into the local energy constraint (\ref{eq: inhom energy constraint}) we obtain the Hamilton-Jacobi equation for $H(\phi)$:
\begin{eqnarray}
\left( \dfrac{\mathrm{d}H}{\mathrm{d}\phi}\right)^2 = \dfrac{3}{2M_{\rm p}^2}H^2 - \dfrac{1}{2M_{\rm p}^4}V(\phi) \label{eq: H-J equation}
\end{eqnarray}
which can be solved for any given potential $V(\phi)$ to give a family of solutions $H(\phi, \mathcal{C})$. One of these solutions combined with:
\begin{eqnarray}
\dfrac{\mathrm{d}\phi}{\mathrm{d}t} = -2M_{\rm p}^2 \,N \, \dfrac{\mathrm{d}H}{\mathrm{d}\phi}
\end{eqnarray} 
and the evolution of the expansion rate (\ref{eq: inhom H expan rate}) offers the complete description of the long wavelength evolution of the inflaton field $\phi$ in the long wavelength metric (\ref{eq:longwavemetric}). \\
\\
It is important at this stage to notice that the naive separate universe picture suggests that at each spatial point we can pick any initial value for the inflaton field and its momentum i.e. that $\mathcal{C} = \mathcal{C}(\textbf{x})$ has an explicit spatial dependence on the initial hypersurface. However this would violate the GR momentum constraint (\ref{eq: GR mom constraint}) which restricts $\mathcal{C}$ to be a \textit{global constant} meaning that all spatial points must be placed along the same integral curve of (\ref{eq: H-J equation}). 
Therefore, once a particular solution of $H(\phi, \mathcal{C})$ of (\ref{eq: H-J equation}) has been obtained the field evolution is given by (for e-fold time $\alpha$):
\begin{eqnarray}
\dfrac{\mathrm{d}\phi}{\mathrm{d}\alpha} &=& -2M_{\rm p}^2 \, \dfrac{\partial \text{ln} H (\phi,\phi_{0})}{\partial \phi} \label{eq: dphidt HJ}
\end{eqnarray}
where $\phi_{0} = M_{\rm p}\sqrt{\frac{2}{3}}~\text{ln}~\mathcal{C}$ is a global constant whose physical significance will become clear later. We see that the inclusion of the GR momentum constraint (\ref{eq: GR mom constraint}) has reduced the dynamics from second order (\ref{eq: inhom K-G}) to first order (\ref{eq: dphidt HJ}) massively reducing the difficulty of the problem. Of particular note this reduction to first order does not make any assumptions about whether the inflaton field is in the slow-roll (or any other) regime. Indeed for slow-roll the GR momentum constraint (\ref{eq: GR mom constraint}) is trivially satisfied but the crucial detail is that the Hamilton-Jacobi equation (\ref{eq: H-J equation}) is valid in any regime, including in the ultra-slow roll regime which is significant for the formation of PBHs.
\subsubsection{Solution for a plateau}
If we imagine that the field enters a plateau region (i.e. $V_{,\phi} = 0$) of the potential of width $\Delta \phi_{pl} \equiv \phi_{in} - \phi_{e}$ from the right hand side with some initial (negative) velocity $\Pi_{in}$\footnote{Entering from the left hand side is equivalent up to a few irrelevant sign changes.}, then the Hamilton-Jacobi (\ref{eq: H-J equation}) can be solved exactly \cite{Prokopec2019}:
\begin{eqnarray}
H(\phi) &= &
\begin{cases}
H_0~\cosh\left( \sqrt{\dfrac{3}{2}} \dfrac{\phi - \phi_0}{M_{\rm p}}\right), & \text{for } \Pi_{in} \neq 0\\[10pt]
H_0 = M_{\rm p}^{-1}\sqrt{\dfrac{V_0}{3}}, & \text{for } \Pi_{in} = 0
\end{cases} \label{eq: H exact flat}
\end{eqnarray}
where $\phi_0$ represents the field value the inflaton asymptotes to. In this sense $\Delta \phi_{cl} \equiv \phi_{in} - \phi_{0}$ represents the distance the classical drift will carry the inflaton as it enters a plateau with finite initial velocity. We can insert the $\Pi_{in} \neq 0$ solution of (\ref{eq: H exact flat}) into the equation of motion (\ref{eq: dphidt HJ}) to find the number of e-folds, $\Delta \alpha$, it takes to reach $\phi$ having started at $\phi_{in}$:
\begin{eqnarray}
\Delta \alpha = -\dfrac{1}{3}\text{ln}\left\lbrace \dfrac{\sinh\left[\sqrt{\frac{3}{2}}\dfrac{\phi-\phi_0}{M_{\rm p}} \right]}{\sinh\left[\sqrt{\frac{3}{2}}\dfrac{\phi_{in}-\phi_0}{M_{\rm p}} \right]}\right\rbrace \label{eq:class e-folds}
\end{eqnarray}
Note however that classically it takes an infinite number of e-folds to reach $\phi_0$ and thus for the classical field velocity $\Pi$ to reach zero. Therefore the $\Pi_{in} = 0$ and  $\Pi_{in} \neq 0$ solutions are completely distinct and there is no way to go between them classically.\\
If we consider the $\Pi_{in} \neq 0$ case then the total distance travelled due to the classical velocity is:
\begin{eqnarray}
\Delta \phi_{cl} \equiv \phi_{in} - \phi_{0} = M_{\rm p} \, {\rm arc}{\sinh} \left(-\dfrac{\Pi_{in}}{\sqrt{2V_0}}\right) = M_{\rm p}\,{\rm arc}{\sinh} \left(  \sqrt{\dfrac{\epsilon_{in}}{3-\epsilon_{in}}} \right) \label{eq: Delta phicl definition}
\end{eqnarray}
where $\epsilon_{in}$ the first Hubble slow-roll parameter -- defined below in equation (\ref{eq: epsilon1 defn}) -- as the field enters the plateau. We see that the range over which the field can slide on the plateau is solely determined by the slow roll parameter associated with the injection velocity. Further assuming slow roll to hold prior to entering the plateau, $\epsilon_{in} \ll 1$, we have
\begin{equation}
\Delta \phi_{cl} \simeq  M_{\rm p} \sqrt{\dfrac{\epsilon_{in}}{3}}
\end{equation}  

\subsection{Stochastic Inflation using the Hamilton-Jacobi equation}
\subsubsection{\texorpdfstring{$\Pi \neq 0$}{}}
If the classical velocity of the field $\Pi \neq 0$ then it follows the Hamilton-Jacobi evolution described by (\ref{eq: H-J equation}). Incorporating the short-wavelength quantum fluctuations results in the addition of a stochastic noise term to (\ref{eq: dphidt HJ}):
\begin{eqnarray}
\dfrac{\mathrm{d}\phi}{\mathrm{d}\alpha} &=& -2 M^2_{\rm p}\,\dfrac{\partial \text{ln} H (\phi,\phi_{0})}{\partial \phi} + \dfrac{H(\phi,\phi_{0})}{2\pi}\xi (\alpha) \label{eq:stochastic H-J} \\
\left\langle \xi (\alpha)\xi (\alpha ')\right\rangle &=& \delta (\alpha - \alpha ') \label{eq: xi defn HJ}
\end{eqnarray}
Where $H (\phi,\phi_{0})$ represents a particular solution to the Hamilton-Jacobi equation (\ref{eq: H-J equation}). If we are in a region of the potential where $V_{,\phi} \neq 0$ or $V_{,\phi} = 0$ but $\phi$ has not yet reached $\phi_{0}$ then $\phi$ still lies on the Hamilton-Jacobi trajectory and equations (\ref{eq:stochastic H-J}) and (\ref{eq: xi defn HJ}) are the appropriate dynamical equations to use.
\subsubsection{\texorpdfstring{$\Pi = 0$}{}}
If the classical velocity $\Pi = 0$ then the field must be on a plateau portion of the potential and have either reached $\phi_{0}$ from a previous Hamilton-Jacobi trajectory or have started in the region with $\Pi = 0$. The problem is therefore equivalent to pure de Sitter with $H = H_{0} = \sqrt{V_{0}/3}$. The field evolution is then simply given by:
\begin{eqnarray}
\dfrac{\mathrm{d}\phi}{\mathrm{d}\alpha} &=& \dfrac{H_{0}}{2\pi}\xi (\alpha) \label{eq: stochastic deSitter}\\
\left\langle \xi (\alpha)\xi (\alpha ')\right\rangle &=& \delta (\alpha - \alpha ') \label{eq: xi defn deSitter}
\end{eqnarray}
We can imagine therefore for a plateau of width $\Delta \phi_{pl} > \Delta \phi_{cl}$ that once $\phi = \phi_{0}$ is reached, the inflaton is injected into the $\Pi= 0$ de Sitter trajectory and freely diffuses along the plateau. What happens at the boundaries is what we cover next.
\subsubsection{Evolution along a finite plateau}
\begin{figure}[tbp]
\centering 
\includegraphics[width=.48\textwidth]{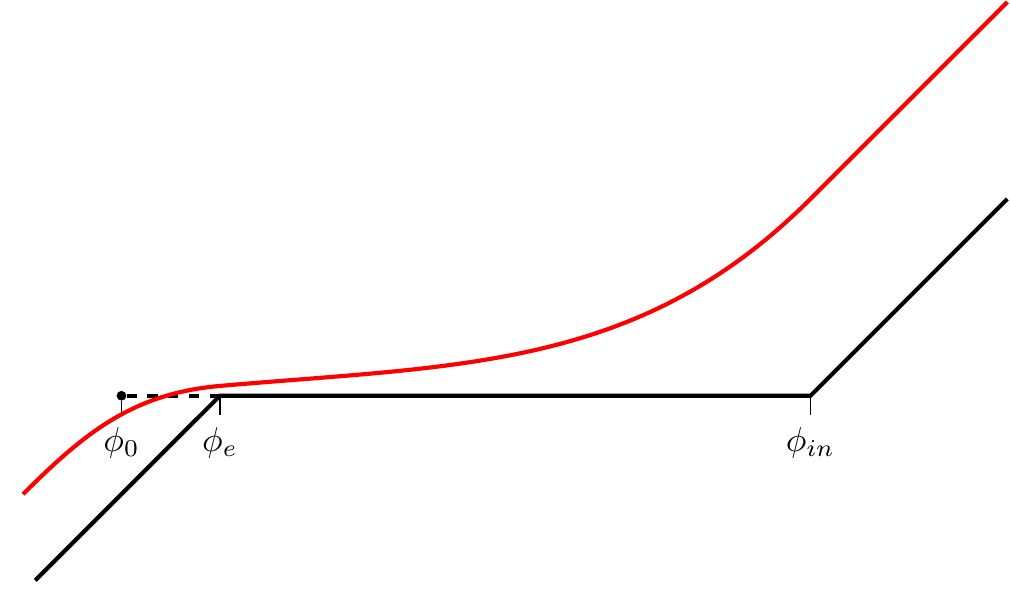}
\includegraphics[width=.48\textwidth]{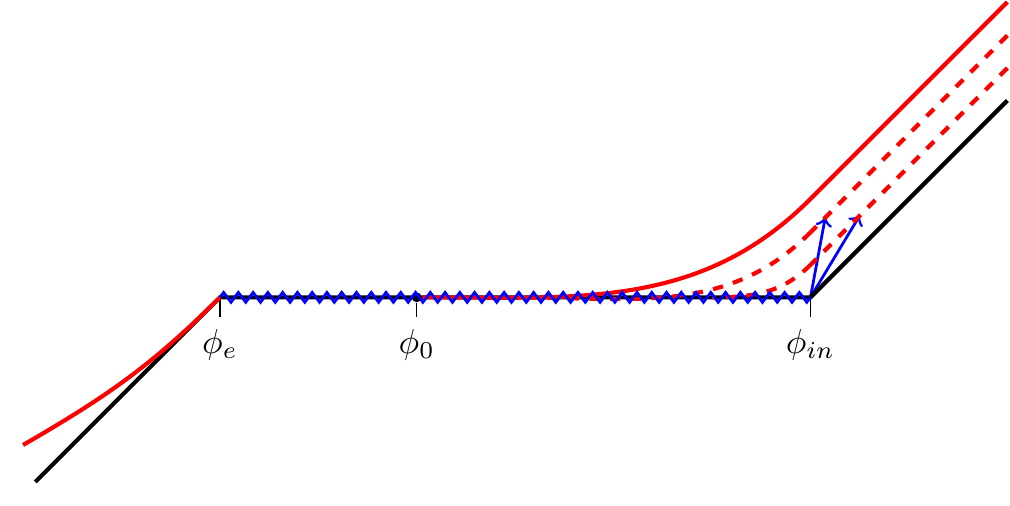}
\caption{\label{fig:Scenario A and B} Scenario A (left panel) corresponds to a plateau shorter than $\Delta \phi_{cl}$, the Hamilton-Jacobi trajectory is plotted in red. Scenario B (right panel) corresponds to a plateau longer than $\Delta \phi_{cl}$, the free diffusion is plotted in blue and the Hamilton-Jacobi trajectory in red as before. When the diffusion reaches $\phi_{in}$ the field jumps onto a new H-J trajectory and will follow it to a new $\phi_0$ before freely diffusing again}
\end{figure}
Consider two scenarios represented in Fig. \ref{fig:Scenario A and B}. Scenario A (left panel) corresponds to a plateau width $\Delta \phi_{pl} < \Delta \phi_{cl}$ i.e. the field enters the region with sufficient velocity to carry it all the way through. The inflaton therefore stays on the Hamilton-Jacobi trajectory for all times. Scenario B (right panel) corresponds to $\Delta \phi_{pl} > \Delta \phi_{cl}$, meaning the inflaton cannot be carried all the way through by classical drift. Therefore, once the inflaton has crossed $\phi_0$ by a stochastic kick it undergoes free diffusion. If the field arrives at the exit point to the plateau, $\phi_{e}$, then the gradient of the potential will start to dominate the evolution and it will have joined a new Hamilton-Jacobi trajectory. If however the field reaches $\phi_{in}$, i.e. the edge of the plateau where it originally entered from, then its evolution is more complicated. The field will jump onto a new H-J curve; the momentum constraint will not be violated when neighbouring spatial points also lie on this new H-J curve and the whole region is surrounded by a zero deterministic velocity boundary. The field will then re-enter the plateau with a different initial velocity, arriving at a new $\phi_0$ before freely diffusing. Scenario B is therefore a highly complicated system to describe. Fortunately -- as we will demonstrate -- realising scenario B is in general forbidden as it leads to an overproduction of PBHs.

\subsection{Stochastic \texorpdfstring{$\delta N$}{} formalism}
As we have described above, the stochastic formulation of inflation allows us to treat the long-wavelength modes of the inflaton, $\phi$, as a classical stochastic variable which obeys stochastic equations of motion -- i.e. equations (\ref{eq:stochastic H-J}) and (\ref{eq: stochastic deSitter}). This means that the time taken (measured in e-folds) for the inflaton to reach a certain point on the potential corresponding to the end of inflation is also a stochastic quantity, denoted by $\mathcal{N}$. We can imagine computing the average e-fold time taken, $\left\langle \mathcal{N}\right\rangle$, by averaging over many different realisations of (\ref{eq:stochastic H-J}) and (\ref{eq: stochastic deSitter}). This is useful because the stochastic $\delta N$ formalism \cite{Enqvist2008, Fujita2013, Fujita2014, Vennin2015} allows one to compute the coarse-grained curvature perturbation on uniform energy density time-slices $\zeta_{cg}$ through:
\begin{eqnarray}
\zeta_{cg} = \mathcal{N} - \left\langle \mathcal{N}\right\rangle
\end{eqnarray}  
which in principle will allow us to compute the PBH mass fraction (\ref{eq:massfracdef}). This can be achieved for example by following the method outlined in \cite{Vennin2015} for slow-roll inflation which uses first passage time analysis on the stochastic differential equation:
\begin{eqnarray}
\dfrac{\mathrm{d}\phi}{\mathrm{d}\alpha} = -\dfrac{v_{,\phi}}{v} + \sqrt{2v}\xi (\alpha) \label{eq:Venin  dphidalpha}
\end{eqnarray} 
where $v \equiv V/24\pi^2M_{Pl}^{4}$ is the dimensionless potential. It is clear that equation (\ref{eq:Venin  dphidalpha}) is of the same form as (\ref{eq: xi defn HJ}) 
\begin{eqnarray}
\dfrac{\mathrm{d}\phi}{\mathrm{d}\alpha} = - \dfrac{\partial_\phi( \tilde{H}^2)}{\tilde{H}^2} + \sqrt{2}\tilde{H}\xi (\alpha) \label{eq:Htilde  dphidalpha}
\end{eqnarray} 
if we make the identification $v \rightarrow \tilde{H}^2$ where $\tilde{H}^2 \equiv H^2/8\pi^2M_{Pl}^{2}$ is the dimensionless Hubble expansion rate and where now $\phi$ is dimensionless . We can therefore utilise the slow-roll formulae given in \cite{Vennin2015} and rewrite them in terms of $\tilde{H}$ giving them full validity outside of the slow-roll regime. For brevity we will simply list the important classical formulae below and list the remainder in Appendix \ref{sec: More Venin formulae}. We first define the first two Hubble slow-roll parameters\footnote{It is worth remembering that despite the name these slow-roll parameters are \textit{exact definitions} and make no a-priori assumption about the inflaton being in a slow-roll regime.}, $\epsilon_1$ \& $\epsilon_2$, in terms of the dimensionless Hubble expansion rate $\tilde{H}$:
\begin{eqnarray}
\epsilon_1 &\equiv & -\dfrac{\mathrm{d}~\text{ln} H}{\mathrm{d}\alpha} = 2M_{Pl}^2\dfrac{\tilde{H}_{,\phi}^2}{\tilde{H}^2} \label{eq: epsilon1 defn} \\
\epsilon_2 &\equiv & -\dfrac{\mathrm{d}~\text{ln} \epsilon_1}{\mathrm{d}\alpha} = 4\dfrac{\tilde{H}_{,\phi\phi}}{\tilde{H}} - \dfrac{2\epsilon_1}{M_{Pl}^2} \label{eq: epsilon2 defn}
\end{eqnarray}
We can then express the \textit{classical}\footnote{By classical we mean that the integrals in Appendix \ref{sec: More Venin formulae} can be well approximated by the leading order contribution in the saddle-point approximation as in \cite{Vennin2015}. The classicality parameter, $\eta_{cl}$, is derived in \cite{Vennin2015} from the second order term in the expansion -- it being small ensures the validity of being in the classical regime. In this sense it is a more sophisticated measure of classicality than simply the ratio of the quantum diffusion over classical drift $\delta \phi_{qu}/\delta \phi_{cl}$ as is often used.} formulae for the average e-fold time, $\left\langle \mathcal{N}\right\rangle$, the deviation from the average e-fold time, $\delta \mathcal{N}^2 = \left\langle \mathcal{N}^2\right\rangle - \left\langle \mathcal{N}\right\rangle^2$, the power spectrum of curvature perturbations, $\mathcal{P}_{\zeta}\vert_{cl}$, and the classicality parameter, $\eta_{cl}$, which must be less than unity for the classical formulae to be good approximations:
\begin{eqnarray}
\left\langle \mathcal{N}\right\rangle\vert_{cl} &=& \dfrac{1}{2}\int_{\phi_{end}}^{\phi}\dfrac{\mathrm{d}x}{M_{Pl}^2}\dfrac{\tilde{H}(x)}{\tilde{H}_{,x}(x)} = \int_{\phi_{end}}^{\phi}\dfrac{\mathrm{d}x}{M_{Pl}}\dfrac{1}{\sqrt{2\epsilon_1}(x)}  \label{eq: classical average e-fold} \\
\delta \mathcal{N}^2\vert_{cl} &=& \dfrac{1}{4}\int_{\phi_{end}}^{\phi}\dfrac{\mathrm{d}x}{M_{Pl}^4}\dfrac{\tilde{H}^{5}(x)}{\tilde{H}_{,x}^{3}(x)} \label{eq: classical var e-fold} \\
\mathcal{P}_{\zeta}\vert_{cl} &=& \dfrac{1}{2}\dfrac{1}{M_{Pl}^2}\dfrac{\tilde{H}^{4}(\phi)}{\tilde{H}_{,\phi}^{2}(\phi)} = \dfrac{\tilde{H}^{2}(\phi)}{\epsilon_{1}} \label{eq: classical power spectrum} \\
\eta_{cl} &=& \left| \dfrac{3}{2}\tilde{H}^2 - \dfrac{\tilde{H}_{,\phi\phi}\tilde{H}^3}{2\tilde{H}_{,\phi}^{2}}\right| = \mathcal{P}_{\zeta}\vert_{cl} \left| \dfrac{7\epsilon_1}{2} - \dfrac{\epsilon_2}{4}\right| \label{eq: classicality criterion}
\end{eqnarray}
We will use these formulae in the next couple of sections to compute the abundance of PBHs.
\section{\label{sec: plateau}Abundance of Primordial Black Holes from a plateau in the potential}
We start by examining how many PBHs are produced in Scenario A where the inflaton's classical velocity when entering the plateau is enough to carry it all the way through, i.e. $\Delta \phi_{pl} \leq \Delta \phi_{cl}$. In \cite{Prokopec2019} it was shown how (\ref{eq:stochastic H-J}) can be reformulated in terms of a Fokker-Planck equation and thus using heat kernel techniques the PDF of e-fold time spent in the plateau $\rho(\mathcal{N})$ can be obtained. Assuming that the inflaton enters the plateau from a previous SR phase at the same e-fold number $N_{in}$ in every stochastic realisation of its trajectory -- see Appendix \ref{sec: SR + USR} where we drop this assumption -- then the PDF $\rho (\mathcal{N})$ for time taken to reach $\phi_{e}$ can be expressed in terms of the difference $\Delta \mathcal{N} \equiv \mathcal{N} - N_{in}$ like so\footnote{Note that the form for $\rho(\mathcal{N})$ shown here is slightly different to the one found in \cite{Prokopec2019} as there was a sign error in their equation (6.4).} \cite{Prokopec2019}:  
\begin{eqnarray}
\rho(\mathcal{N}) =& & \dfrac{3}{\sqrt{\pi}}\chi\sqrt{n+1}\left[ (n+1)e^{-3\Delta \mathcal{N}} -n\sigma \right]\text{exp}\left[ -(n + 1)\chi^2(\sigma -e^{-3\Delta \mathcal{N}})^2 \right]  \nonumber \\
&+& \dfrac{3}{\sqrt{\pi}}\chi\sqrt{n}\left[ n + 1 + \sigma (n + 2)e^{-3\Delta \mathcal{N}} \right]e^{Y}\text{exp}\left[ -n\chi^2(1 + \sigma e^{-3\Delta \mathcal{N}})^2 \right]  \nonumber \\
&-& 6 \chi^2\sigma \left[ \sigma e^{-3\Delta \mathcal{N}} - n(2n + 3)\right]e^{Y}e^{-3\Delta \mathcal{N}}\text{erfc}\left[\sqrt{n}\chi(1 + \sigma e^{-3\Delta \mathcal{N}})\right] \label{eq:rhoN HJ} \\
Y \equiv &-& \chi^2\sigma\left[4n e^{-3\Delta \mathcal{N}} -\sigma (1 + 2e^{-6\Delta \mathcal{N}})  \right] \label{eq: Y defn} \\
n \equiv & & \dfrac{1}{e^{6\Delta \mathcal{N}} - 1}
\end{eqnarray}
where we introduced the dimensionless parameter $\chi$:
\begin{eqnarray}
\chi \equiv  \sqrt{\dfrac{3}{2v_0}}\dfrac{\Delta \phi_{cl}}{M_{pl}} %= \sqrt{\dfrac{3}{2v_0}}\sqrt{\dfrac{\epsilon_{in}}{3-\epsilon_{in}}}
\simeq\sqrt{\dfrac{\epsilon_{in}}{2v_0}} \label{eq: chi_in defn}
\end{eqnarray}
defined in terms of the classical drift distance, $\Delta \phi_{cl}$ - equivalently the slow roll parameter $\epsilon_{1}$ of the prior SR region - and the dimensionless plateau height $v_0 = V_0/24\pi^2M_{pl}^{4}$. We also introduced the dimensionless parameter $\sigma$:
\begin{eqnarray}
\sigma \equiv \dfrac{\Delta \phi_{cl} - \Delta \phi_{pl} }{\Delta \phi_{cl}} \label{eq:sigma defn}
\end{eqnarray}
which parametrises how wide the plateau is relative to the classical drift distance $\Delta \phi_{cl}$. Notice that $\sigma = 0$ corresponds to $\Delta \phi_{cl} = \Delta \phi_{pl}$ and that the limit $\sigma \rightarrow 1$ corresponds to $\Delta \phi_{pl} \rightarrow 0$. We have also introduced erfc$(x)$ which is the standard complementary error function.\\

\subsection{The case \texorpdfstring{$\Delta \phi_{pl} < \Delta \phi_{cl}$}{} ($0 < \sigma <1$)}

\begin{figure}[tbp]
\centering 
\includegraphics[width=.75\textwidth]{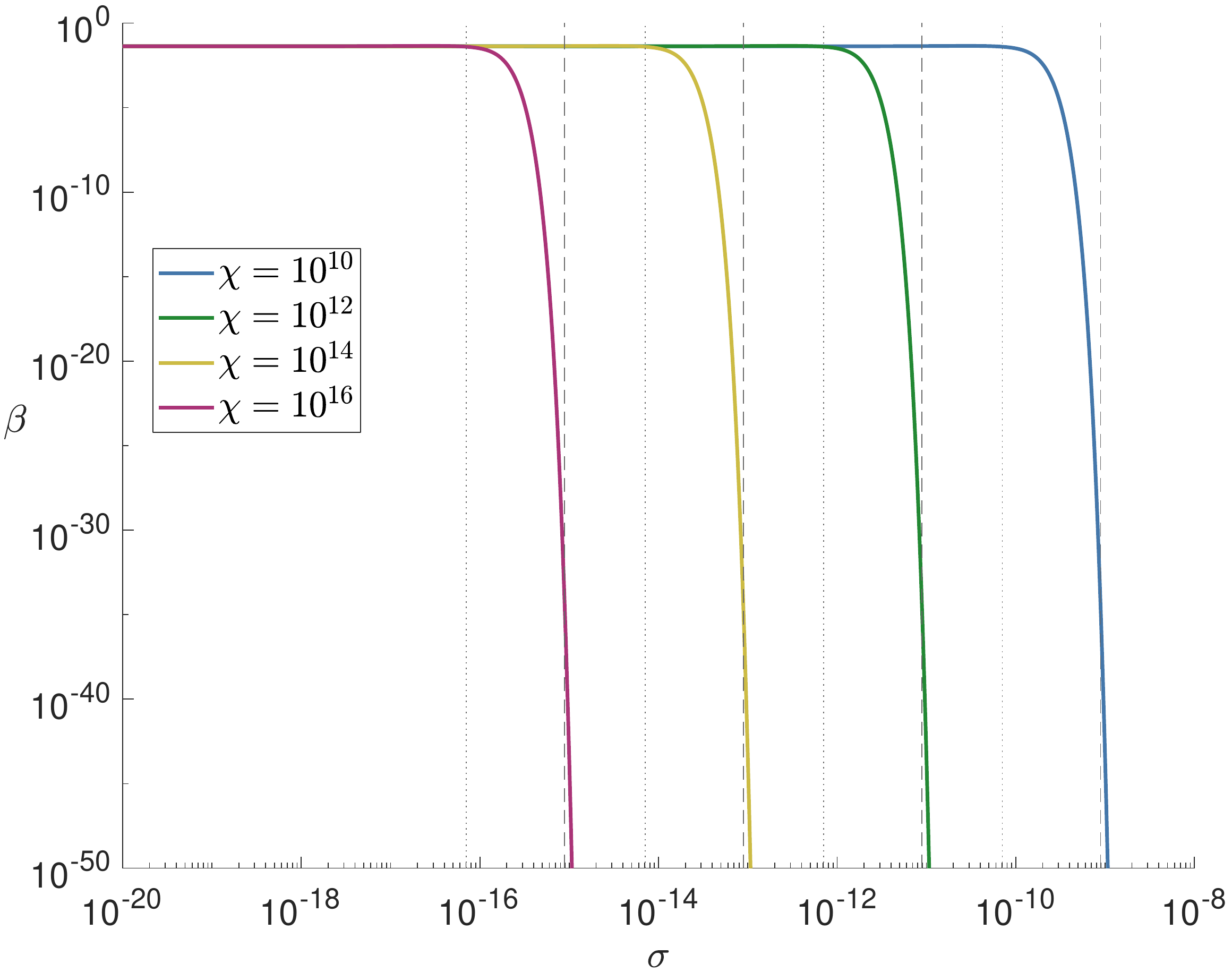}
\caption{\label{fig:chi_sigma_forBeta_USRandSR} The mass fraction of PBHs, $\beta$, as a function of $\sigma$ for four values of $\chi$ computed using (\ref{eq: Beta chie > 0}). The dotted lines represent the scale when the classical approximation fails, $\sigma_{cl}$, given by (\ref{eq:sigma classical}). The dashed lines represent the pivot scale to overproduce PBHs, $\sigma_p$, predicted by (\ref{eq:sigma pivot}). It is clear that values of $\sigma$ significantly smaller than $\sigma_p$ overproduced PBHs whereas values of $\sigma$ larger than this value correspond to negligible production.}
\end{figure}

If we consider the large $\Delta \mathcal{N}$ limit i.e. deep in the tail where the PBH mass fraction is calculated then (\ref{eq:rhoN HJ}) simplifies to:
\begin{eqnarray}
\rho (\mathcal{N}) \simeq \dfrac{6}{\sqrt{\pi}}\chi ~e^{-\chi^{2}\sigma^2}e^{-3\Delta \mathcal{N}} \label{eq: rhoN HJ LARGE N}
\end{eqnarray}
Using the mass fraction definition (\ref{eq:massfracdef}) under the large $\Delta \mathcal{N}$ limit (\ref{eq: rhoN HJ LARGE N}) we obtain:
\begin{eqnarray}
\beta (M) \simeq \dfrac{4}{\sqrt{\pi}}\chi ~e^{-\chi^{2}\sigma^2}~e^{-3(\zeta_c + \left\langle \mathcal{N}\right\rangle - N_{in})}\underbrace{e^{\frac{9}{4}\chi^{-2}}}_{SR} \label{eq: Beta chie > 0}
\end{eqnarray} 
where we have also included the contribution from a previous slow-roll phase -- see Appendix \ref{sec: SR + USR}. This slow-roll contribution is only important for $\chi \ll 1$ where it very quickly forces $\beta$ to unrealistically large values. The exponential tail of (\ref{eq: rhoN HJ LARGE N}) means that the mass fraction does not depend on the combination $\zeta_c /\delta \zeta$ where $\delta \zeta$ is the variance of the perturbations. To see how the mass fraction does depend on the variance we note that the classical power spectrum given in (\ref{eq: classical power spectrum}) in this case is given by $\mathcal{P}_{\zeta}\vert_{cl} \sim 1/ 3\chi^2\sigma^2$ suggesting that the variance of perturbations is given by $\delta \zeta \sim 1/\chi^2\sigma^2$ while the evolution is classically dominated. 

Looking at equation (\ref{eq: Beta chie > 0}) it is clear for large values of $\chi$ that the $e^{-\chi^{2}\sigma^2}$ factor forces $\beta$ to be incredibly small unless $\sigma$ is very small. Assuming we can approximate the average e-fold time by its classical value $\left\langle \mathcal{N}\right\rangle \simeq \left\langle \mathcal{N}\right\rangle\vert_{cl}$ which can be computed from (\ref{eq: classical average e-fold}) as:
\begin{eqnarray}
\left\langle \mathcal{N}\right\rangle\vert_{cl} \simeq -\dfrac{1}{3}\text{ln}(\sigma)
\end{eqnarray}
This allows us to rewrite (\ref{eq: Beta chie > 0}) as:
\begin{eqnarray}
\beta (M) \simeq \dfrac{4}{\sqrt{\pi}}\chi \sigma ~e^{-\chi^{2} \sigma^{2}}~e^{-3\zeta_c} \simeq \dfrac{4}{\sqrt{\pi}}\chi\sigma ~e^{-3\zeta_c} \label{eq: Beta chie > 0 small sigma}
\end{eqnarray}
where the second approximation uses the fact that $x~e^{-x^2} \sim x$ for small $x$.
We can use (\ref{eq: Beta chie > 0 small sigma}) to define a scale for $\sigma$ where the mass fraction is important. We therefore find that:
\begin{eqnarray}
\sigma &\ll &\dfrac{\sqrt{\pi}}{4\chi}~e^{3\zeta_c} \Leftrightarrow \beta \text{ Violates constraints} \nonumber\\
\sigma &\gg &\dfrac{\sqrt{\pi}}{4\chi}~e^{3\zeta_c} \Leftrightarrow \beta \text{ Negligible} \nonumber \\ \label{eq:sigma pivot}
\end{eqnarray}
and we therefore identify a pivot scale by 
\begin{equation}
\sigma_{p} \equiv \dfrac{\sqrt{\pi}}{4\chi}~e^{3\zeta_c}
\end{equation}
This scale is verified in Fig.~\ref{fig:chi_sigma_forBeta_USRandSR} where we plot, using (\ref{eq: Beta chie > 0}),  the dependence of the mass fraction, $\beta$ on $\sigma$ for a few values of $\chi$. We can see that the behaviour of $\beta$ almost looks like a step function with a sharp drop off as $\sigma$ is increased.  The dashed lines -- corresponding to the scale predicted by (\ref{eq:sigma pivot}) -- accurately describes where this sharp dropoff takes place and marks the separation between overproduction of PBHs and negligible production. 

We can verify our use of $\left\langle \mathcal{N}\right\rangle \simeq \left\langle \mathcal{N}\right\rangle\vert_{cl}$ by computing the value of $\sigma$ for which the classicality parameter is violated, $\eta_{cl}=1$. Using equation (\ref{eq: classicality criterion}) we find that the classicality parameter evaluated at $\phi_e$ is given by:
\begin{eqnarray}
\eta_{cl}(\phi_{e}) \simeq \left| \dfrac{3}{2}\tilde{H}_{0}^{2} - \dfrac{1}{2\chi^2\sigma^2}\right| \simeq \dfrac{1}{2\chi^2\sigma^2} \label{eq:etacl for sig > 0}
\end{eqnarray}
which suggests that smaller (bigger) values of the combination $\chi^2\sigma^2$ correspond to being in the quantum (classical) regime. This also justifies the use of the second approximation in (\ref{eq: Beta chie > 0 small sigma}) as the exponential dependence on $\chi^2\sigma^2$ which massively suppresses the formation of PBHs also corresponds to being deep in the classical regime. This transition from classically dominated to quantum diffusion dominated dynamics takes place when $\eta_{cl}=1$ which we substitution into (\ref{eq:etacl for sig > 0}) to identify this transition:
\begin{eqnarray}
\sigma & < &\dfrac{1}{\sqrt{2}\chi} \Leftrightarrow \text{ The inflaton dynamics has a diffusion dominated regime} \nonumber\\
\sigma & > &\dfrac{1}{\sqrt{2}\chi} \Leftrightarrow \text{ Inflaton evolution is always classically dominated} \nonumber \\ \label{eq:sigma classical}
\end{eqnarray}
with $\eta_{cl} =1$ at $\sigma_{cl} = 1/\sqrt{2}\chi$. As $\sigma_{cl} < \sigma_{p}$, we are consistent in using $\left\langle \mathcal{N}\right\rangle\vert_{cl}$ to evaluate $\sigma_p$. However,  this result has a more significant consequence. As can be clearly shown by the dotted lines in Fig.~\ref{fig:chi_sigma_forBeta_USRandSR} one only enters the diffusion dominated regime once the mass fraction of PBHs, $\beta$, is prohibitively high. This value is given by substituting $\sigma_{cl}$ into $\beta$:
\begin{eqnarray}
\beta_{\sigma_{cl}} \sim \dfrac{4}{\sqrt{2\pi e}}e^{-3\zeta_c} \sim 0.4289 \cdot  \dfrac{4}{\sqrt{\pi}}e^{-3\zeta_c} \label{eq:beta_plat_classical}
\end{eqnarray}
In other words the classically dominated evolution will already overproduce PBHs before the inflaton even enters the diffusion dominated regime. We will expand on this point in the next subsection. 

\subsection{The case \texorpdfstring{$\Delta \phi_{pl} = \Delta \phi_{cl}$ ($\sigma=0$) }{}}

\begin{figure}[tbp]
\centering 
\includegraphics[width=.48\textwidth]{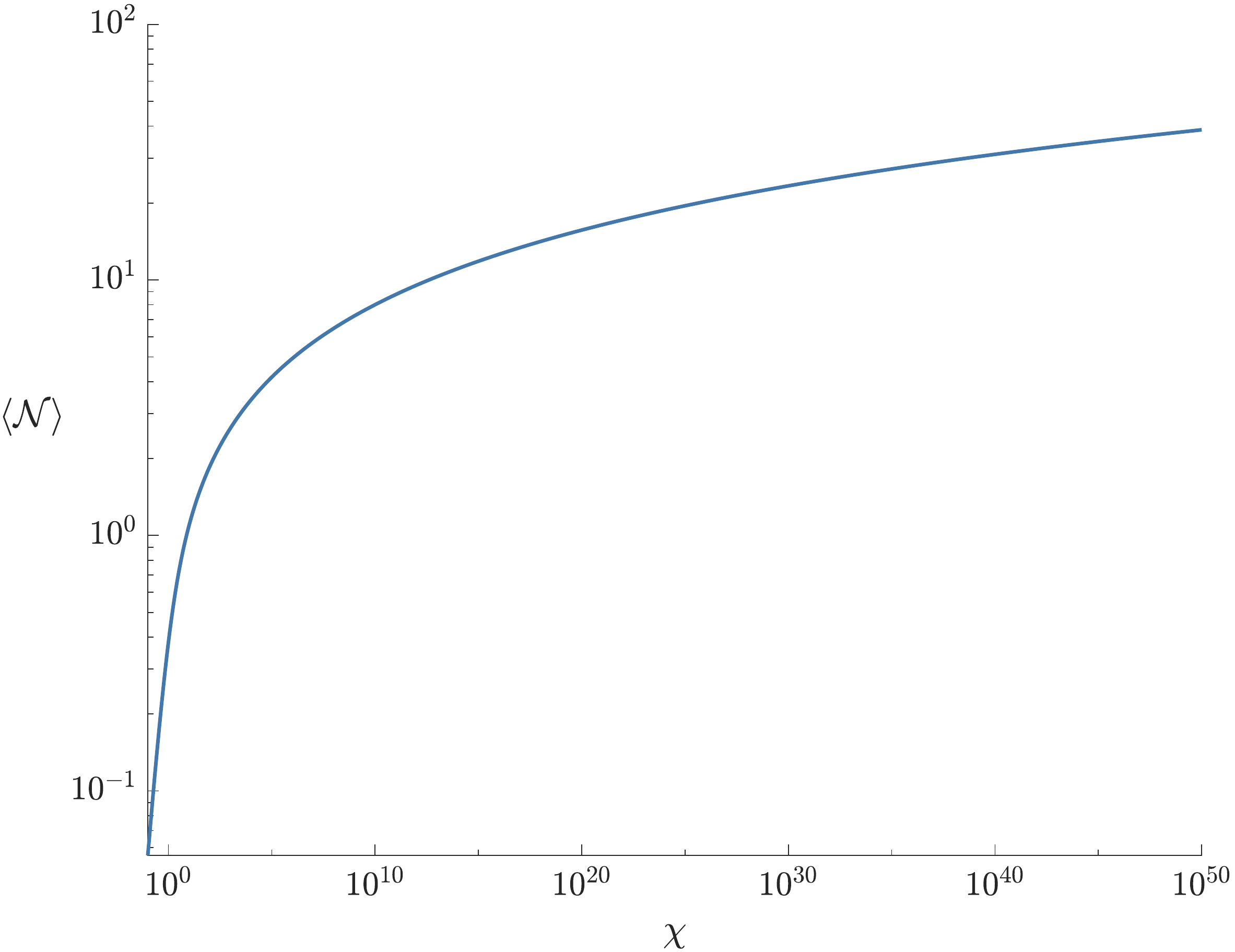}
\includegraphics[width=.48\textwidth]{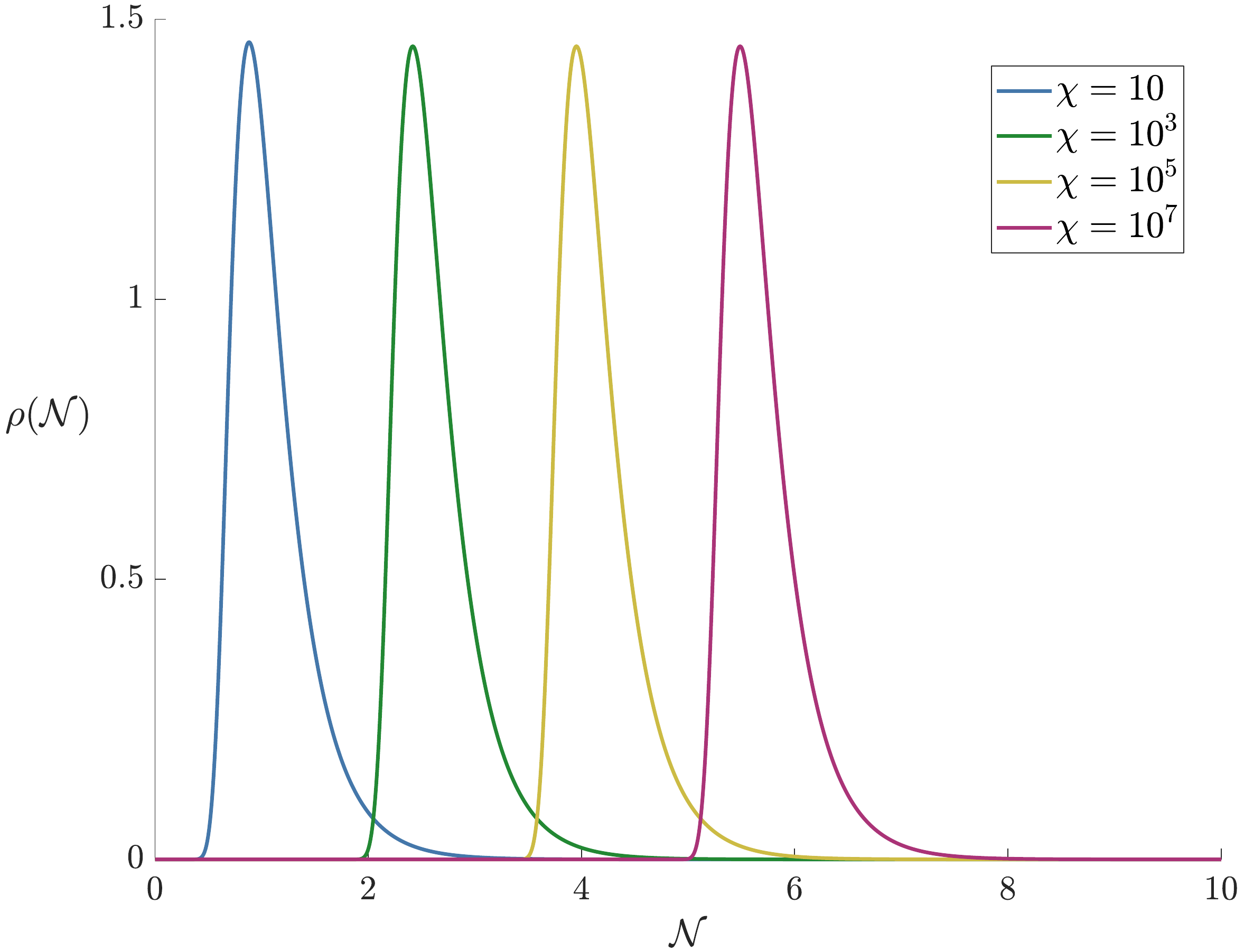}
\caption{\label{fig:rhoN_chi_e_zero.pdf} The dependence of average number of e-folds $\left\langle \mathcal{N}\right\rangle$ realised in the quantum well on $\chi$ (left) and the PDF $\rho (\mathcal{N})$ of e-fold time spent in the plateau for three values of $\chi$ (right). Both for $\Delta \phi_{cl} = \Delta \phi_{pl}$.}
\end{figure}
\begin{figure}[tbp]
\centering 
\includegraphics[width=.75\textwidth]{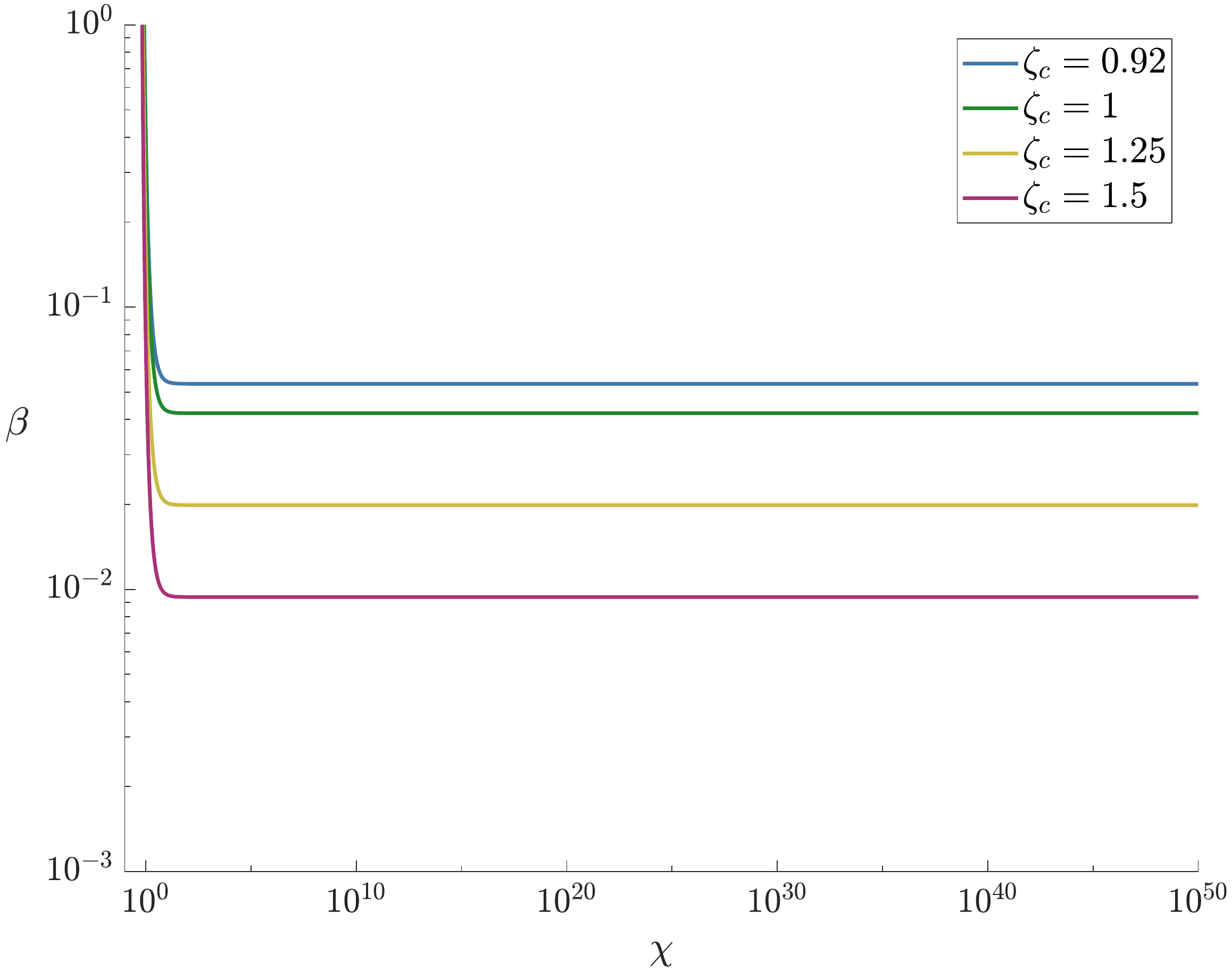}
\caption{\label{fig:Beta_USR_chi_e_zero.pdf} Dependence of mass fraction $\beta$ on $\chi$ using (\ref{eq: Massfrac chie = 0}) for four different values of the cutoff $\zeta_c$ for  $\Delta \phi_{cl} = \Delta \phi_{pl}$. The sudden spike in $\beta$ for small values of $\chi$ is the contribution from the previous SR phase}
\end{figure}
%\begin{figure}[tbp]
%\centering 
%\includegraphics[width=.48\textwidth]{fig/Betaeps_USR_chi_e_zero.pdf}
%\includegraphics[width=.48\textwidth]{fig/v0epsilon_forBeta_USR_chi_e_zero.pdf}
%\caption{\label{fig:v0eps constraints chie = 0} The mass fraction $\beta$ as a function of $\epsilon_{in}$ as the inflaton enters the plateau for four different plateau heights corresponding to $v_0$  (left) and necessary values of  $\epsilon_{in}$ and $v_0$ to satisfy the upper bounds on $\beta$ for three different PBH mass ranges (right) -- in both cases $\Delta \phi_{pl} = \Delta \phi_{cl}$.}
%\end{figure}

In the limit $\Delta \phi_{pl} = \Delta \phi_{cl}$ (equivalently $\sigma = 0$), (\ref{eq:rhoN HJ}) simplifies to:
\begin{eqnarray}
\rho(\mathcal{N}) = \dfrac{6}{\sqrt{\pi}} \sqrt{n}(n+1)\chi ~e^{-n\chi^{2}}  \label{eq:rho chie = 0}
\end{eqnarray}
If we consider the large $\Delta \mathcal{N}$ limit of (\ref{eq:rho chie = 0}) to examine the behaviour in the tail we obtain:
\begin{eqnarray}
\rho(\mathcal{N})\sim \dfrac{6}{\sqrt{\pi}}\chi ~e^{-3\Delta\mathcal{N}} \label{eq:rho chie = 0 approx}
\end{eqnarray}
which corresponds to (\ref{eq: rhoN HJ LARGE N}) in the $\sigma \rightarrow 0$ limit as it should. Importantly therefore it also exhibits the same non-gaussian exponential tail $e^{-3\Delta\mathcal{N}}$. The average number of e-folds $\left\langle \mathcal{N}\right\rangle$ realised can be computed exactly from (\ref{eq:rho chie = 0}) and is given by:
\begin{eqnarray}
\left\langle \mathcal{N}\right\rangle = \dfrac{\pi}{6}\text{erfi}(\chi)-\dfrac{\chi^{2}}{3}\,\, {}_2F_2 \left( \lbrace 1,1\rbrace, \left\lbrace \dfrac{3}{2},2\right\rbrace ,\chi^{2}\right) 
\end{eqnarray}
where erfi is the imaginary error function and $_2F_2$ is a generalized hypergeometric function. Note in practice that for very large values of $\chi$ it is usually more practical numerically to compute $\left\langle \mathcal{N}\right\rangle$ directly from the PDF.

In Fig.~\ref{fig:rhoN_chi_e_zero.pdf} we plot both the dependence of average e-fold time spent in the plateau, $\left\langle \mathcal{N}\right\rangle$, on $\chi$ (left panel) and the PDF $\rho(\mathcal{N})$ for four values of $\chi$ (right panel). In the left panel we see how the average number of e-folds $\left\langle \mathcal{N}\right\rangle $ grows with $\chi$ very quickly initially before growing logarithmically at very large values of $\chi$. At $\chi \sim 10^{50}$ the average time spent in the plateau, $\left\langle \mathcal{N}\right\rangle $, is comparable to the total duration of inflation.\footnote{Strictly speaking we mean the average time spent in the plateau, $\left\langle \mathcal{N}\right\rangle $, is longer than the allowed number of e-folds between CMB modes exiting the horizon and inflation ending.} 
Looking at the right panel we can see how $\rho(\mathcal{N})$ is non-gaussian with a deep tail and that this shape does not change noticeably as $\chi$ is increased by several orders of magnitude. Indeed, the only noticeable impact of  increasing $\chi$ is to translate the whole PDF to the right. This suggests that even if the inflaton spends a large amount of time on the plateau it is still reasonably localised in time around its average value and we can reasonably assign a time for these modes to exit the horizon. 

The PBH mass fraction for the exact PDF (\ref{eq:rho chie = 0}) can be calculated exactly as:
\begin{eqnarray}
\beta (M) = 2~\text{erf}(\sqrt{n_{c}}\chi)\times \underbrace{e^{\frac{9}{4}\chi^{-2}}}_{\text{SR}}  \label{eq: Massfrac chie = 0}
\end{eqnarray}
where
\begin{eqnarray}
n_{c} \equiv \dfrac{1}{e^{6(\zeta_c + \left\langle \mathcal{N}\right\rangle - N_{in})} -1} \label{eq:nc defn}
\end{eqnarray}
and again we have included the contribution from the previous SR phase.
Note that if we consider the large $\Delta \mathcal{N}$ limit of (\ref{eq: Massfrac chie = 0}) then it reduces to:
\begin{eqnarray}
\beta (M) \sim \dfrac{4}{\sqrt{\pi}}\chi~e^{-3(\zeta_c + \left\langle \mathcal{N}\right\rangle - N_{in})}\times e^{\frac{9}{4}\chi^{-2}}  \label{eq: massfrac chie = 0 approx}
\end{eqnarray}
We see that this is the $\sigma \rightarrow 0$ limit of equation (\ref{eq: Beta chie > 0}) confirming the two results are consistent with each other. It is worth appreciating that, like in the analysis of \cite{Pattison2017}, the PDF $\rho(\mathcal{N})$, average number of e-folds $\left\langle \mathcal{N}\right\rangle$ and mass fraction of PBHs $\beta$ only depends on a single parameter $\chi$ \footnote{In \cite{Pattison2017} their parameter is $\mu \equiv \Delta \phi_{pl}/\sqrt{v_0}$ which for $\sigma = 0$ is related to $\chi$ through $\chi = \mu\sqrt{3/2}$.}. However unlike in \cite{Pattison2017} this computation fully takes into account the velocity of the inflaton as it enters the plateau albeit in the restricted case where $\Delta \phi_{pl} = \Delta \phi_{cl}$. \\

In Fig.~\ref{fig:Beta_USR_chi_e_zero.pdf} we plot the dependence of the mass fraction, $\beta$, on $\chi$ for four different values of the cutoff, $\zeta_c$, between the lower and upper limits of $\sim$ 0.92 and 1.5 permitted \cite{Musco2019}. We can see that -- apart from the sharp spike at small $\chi$ due to the previous SR phase -- the mass fraction is constant for all values of $\chi$ and approximately lies in the range $10^{-1}$ - $10^{-2}$. The mass fraction, $\beta$, is therefore in excess of all the upper limits imposed in the possible mass ranges for PBHs.\\

This constant value occurs because for large $\chi$ the quantity $\chi ~e^{-3(\left\langle \mathcal{N}\right\rangle -N_{in})} \approx 0.3747$. We can therefore say that the mass fraction converges very quickly for $\chi \gg 1$ to:
\begin{eqnarray}
\beta (M) \sim 0.3747\times\dfrac{4}{\sqrt{\pi}}~e^{-3\zeta_c} \label{eq: massfrac chie = 0 const approx}
\end{eqnarray}
This equation very accurately describes the horizontal lines displayed in Fig.~\ref{fig:Beta_USR_chi_e_zero.pdf}.\\

It is therefore clear from the analysis of this section that for any plateau of equal width to the classical drift distance that one will generically overproduce PBHs for any remotely realistic inflationary potential. This means that Scenario B in Fig.~\ref{fig:Scenario A and B} is completely ruled out as a subsequent phase of free diffusion would only \textit{enhance} the curvature perturbation, producing even more PBHs; we verify this in the following subsection. Not only that but as shown in Fig.~\ref{fig:chi_sigma_forBeta_USRandSR} even a diffusion dominated regime with non-zero classical drift is forbidden. We therefore arrive at the main result of this section:  \\

\textit{Any period of quantum diffusion dominated dynamics on a plateau will overproduce PBHs}\\

In the next section we will relax the assumption of a perfectly level plateau to local inflection points. Again, we will find that diffusion domination overproduces PBHs, demonstrating that this result is generic for such inflationary potentials. Before doing so however, we verify that making the plateau even longer makes things worse for the overproduction of PBHs, as expected.     

\subsection{The case \texorpdfstring{$\Delta \phi_{pl} > \Delta \phi_{cl}$}{} ($ \sigma < 0$)}
\begin{figure}[tbp]
\centering 
\includegraphics[width=.48\textwidth]{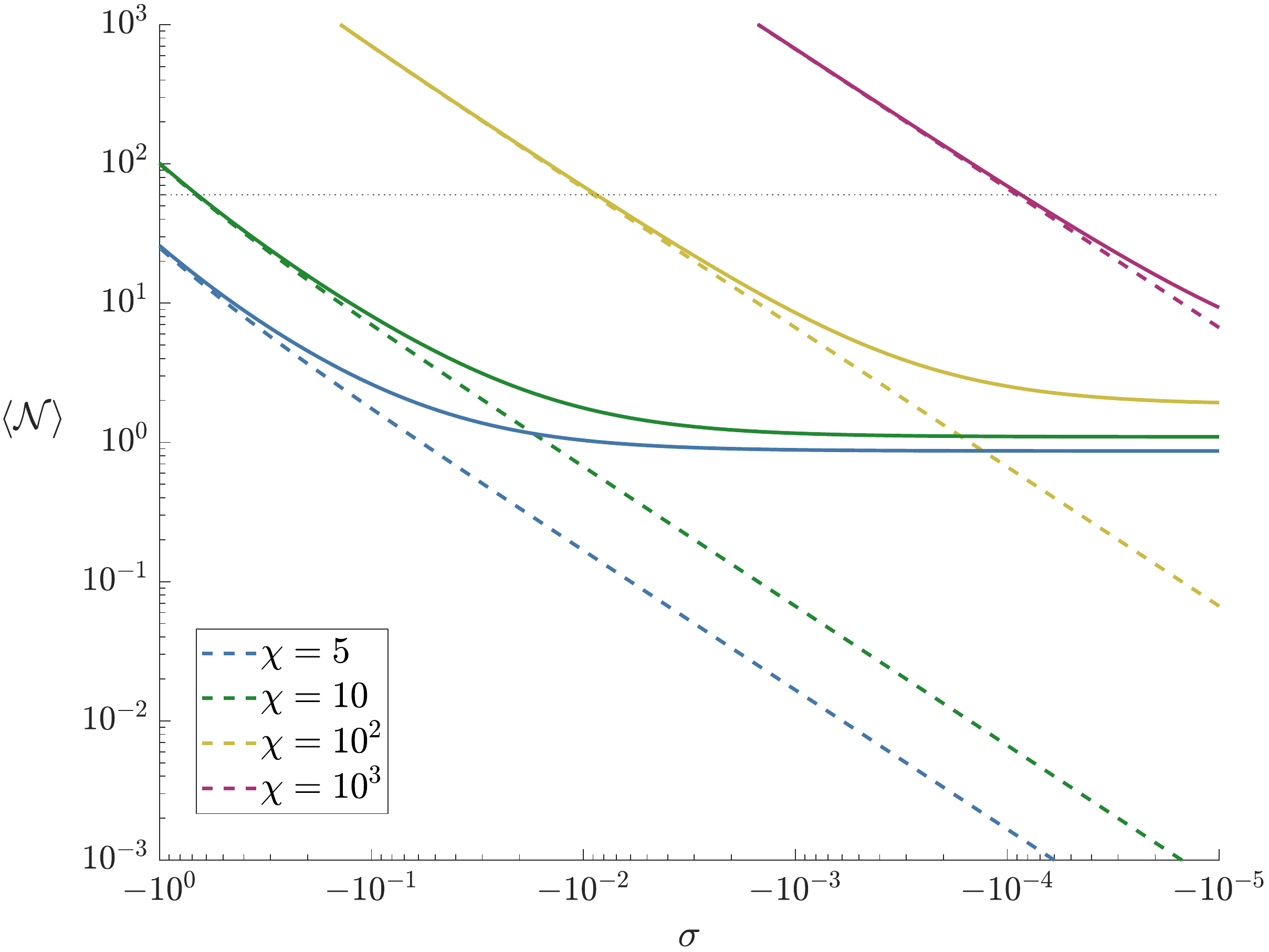}
\includegraphics[width=.48\textwidth]{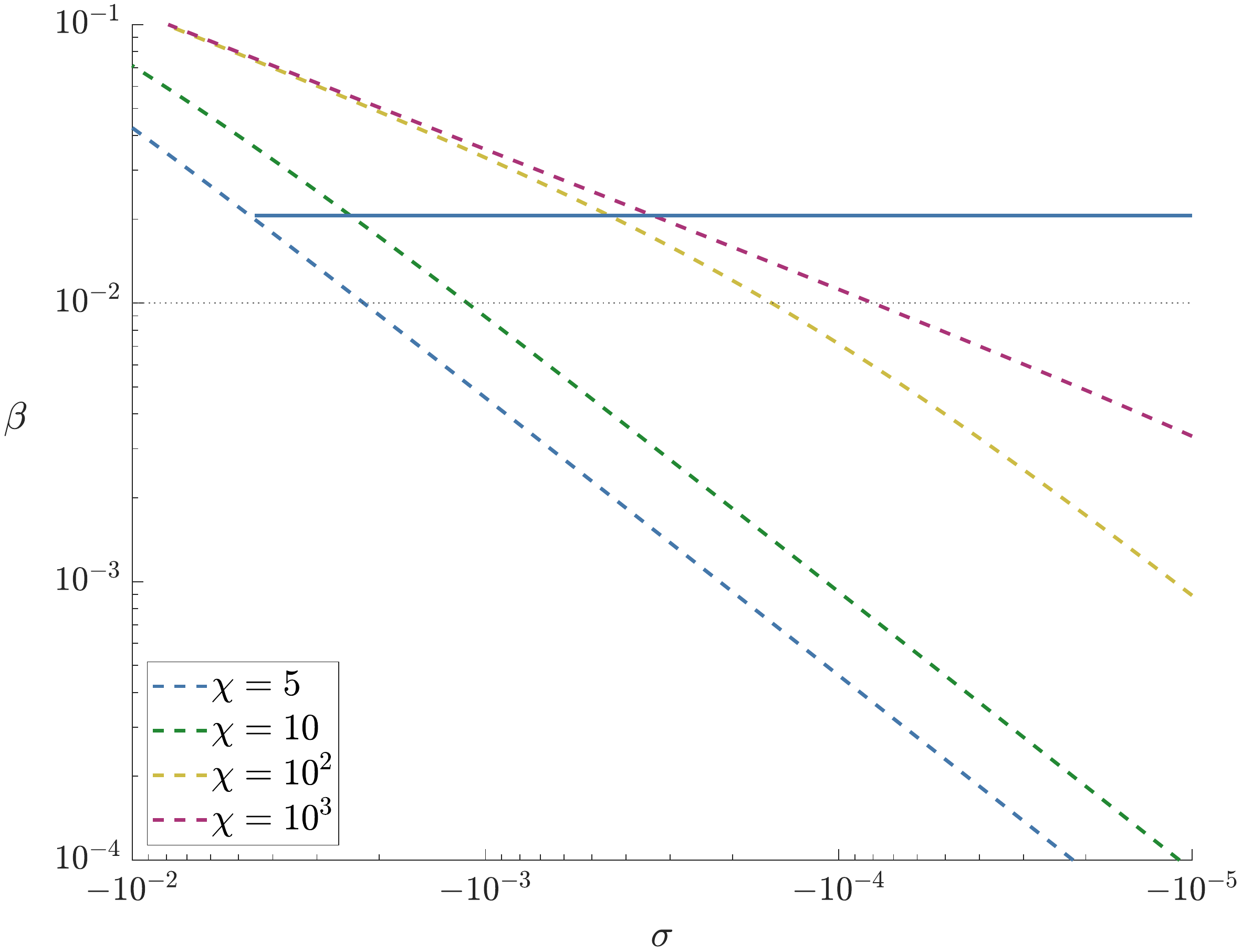}
\caption{\label{fig:Pattison muN and beta} The dependence on $\sigma$ of the average number of e-folds $\left\langle \mathcal{N}\right\rangle$ (left) and the mass fraction $\beta$ (right) given as a function of $\sigma<0$ ($\Delta \phi_{pl} > \Delta \phi_{cl}$). The dashed lines correspond to the free diffusion value only and the solid lines to the total of the H-J plus free diffusion phase. For small $\sigma$, $\left\langle \mathcal{N}\right\rangle$ in the H-J computation levels off to the number of e-folds required for the field to slide on the plateau via its initial velocity with free diffusion making a very small contribution. The horizontal dotted line on the left plot corresponds to 60 e-folds. The mass fraction accounting for both phases given by (\ref{eq:rhoHJ+dS}) is shown for $\chi = 5$ by the solid blue line in the right plot where the horizontal dotted line corresponds to the weakest ($10^{-2}$) bound on PBH abundance.}
\end{figure}
If the plateau is wider than the classical drift distance, the field enters a period of free diffusion as described by scenario B in Fig.~\ref{fig:Scenario A and B}. When the field reaches $\phi_0$ it exits the H-J trajectory and enters a period of free diffusion with zero drift velocity. If we assume -- for now -- that the distribution enters as a delta function then we can use the results of Pattison \textit{et.~al} \cite{Pattison2017} to describe this second phase. The PDF for exit time, $\rho_{dS}$, average time spent during free diffusion, $\left\langle  \mathcal{N}\right\rangle_{dS}$, and the mass fraction, $\beta_{dS}$, for this pure de Sitter phase are given by \cite{Pattison2017}:
\begin{eqnarray}
\rho_{dS}(\mathcal{N}) &=& \dfrac{3\pi}{\chi^2 (1-\sigma)^2}\sum_{n=0}^{\infty}\left( n+ \dfrac{1}{2} \right)\text{sin}\left[ \dfrac{\sigma}{\sigma-1}\pi\left( n+ \dfrac{1}{2}\right)\right]\text{exp}\left[ -\dfrac{3\pi^2}{2\chi^2 (1-\sigma)^2}\left( n+ \dfrac{1}{2}\right)^2 \mathcal{N}\right]\nonumber \\\label{eq:PDF de Sitter}\\
\left\langle \mathcal{N} \right\rangle_{dS} &=& \dfrac{2\chi^2}{3}\sigma \left( \dfrac{\sigma}{2} -1\right)\label{eq:muN dS}\\
\beta_{dS} &=& \dfrac{4}{\pi}\sum_{n=0}^{\infty}\dfrac{\text{sin}\left[ \dfrac{\sigma}{\sigma-1}\pi\left( n+ \dfrac{1}{2}\right)\right]}{ n+ \dfrac{1}{2}}\text{exp}\left[ -\dfrac{3\pi^2}{2\chi^2 (1-\sigma)^2}\left( n+ \dfrac{1}{2}\right)^2 (\zeta_c + \left\langle \mathcal{N}_3 \right\rangle )\right]\nonumber \\\label{eq:massfrac de Sitter}
\end{eqnarray}
expressed in terms of our parameters $\chi$ and $\sigma$. $\left\langle \mathcal{N} \right\rangle_{dS}$ and $\beta_{dS}$ are plotted as dashed lines on the left and right plots of Fig.~\ref{fig:Pattison muN and beta} respectively as functions of $\sigma$ for different values of $\chi$. We see that unless $\sigma$ is very small in absolute value, the average number of e-folds realised throughout the plateau can easily exceed the number of e-folds needed between the CMB and the end of inflation, at least for inflationary scales not too close to $M_p$ (i.e. for large $\chi$). The mass fraction given by (\ref{eq:massfrac de Sitter}) places tighter bounds on $\sigma$, forcing it to be small in absolute value to not violate constraints. All these conclusion are drawn for the pure diffusion computation.

However, these computations have assumed that the field starts the free diffusion phase at a fixed time and localized on the plateau. This is clearly not true as is evident e.g. from the right plot of Fig.~\ref{fig:rhoN_chi_e_zero.pdf} where a similar looking PDF would determine the starting time of the post-H-J free diffusion phase. To include the effect on the mass fraction of the prior slide of the field on the plateau, one should instead do the convolution of the H-J phase (\ref{eq:rhoN HJ}) and the free diffusion phase (\ref{eq:PDF de Sitter}), and then compute the mass fraction from this convoluted PDF. While in principle this can be done numerically in a similar way as the procedure outlined in appendix \ref{sec: SR + USR} we find that the more illuminating method is to modify the procedure presented in \cite{Prokopec2019} to account for a finite width plateau. The full procedure is outlined in appendix \ref{sec: PDF for scenario B} but the main result is the PDF for exit times, $\rho_{HJ + dS} (\mathcal{N})$, given by:
\begin{eqnarray}
\rho_{HJ+dS}(\mathcal{N}) &=& \dfrac{2\sqrt{\pi}}{\chi(1-\sigma)^2}\int_{0}^{\mathcal{N}}\mathrm{d}u~\left[\text{sinh}(3u)\right]^{-3/2}~\text{exp}\left[\dfrac{3}{2}u -\dfrac{1}{2}\left( \text{coth}(3u ) -1\right)\chi^2\right] \nonumber \\
&&\times \sum_{n=0}^{\infty}\left(n+\dfrac{1}{2}\right)\text{sin}\left[\left((n+\dfrac{1}{2}\right)\dfrac{\pi\sigma}{\sigma-1}\right]\text{exp}\left[-\left(n + \dfrac{1}{2}\right)^2\dfrac{2\pi^2}{3\chi^2 (1-\sigma)^2} (\mathcal{N} -u) \right] \nonumber \\ \label{eq:rhoHJ+dS}
\end{eqnarray}

\begin{figure}[tbp]
\centering 
\includegraphics[width=.48\textwidth]{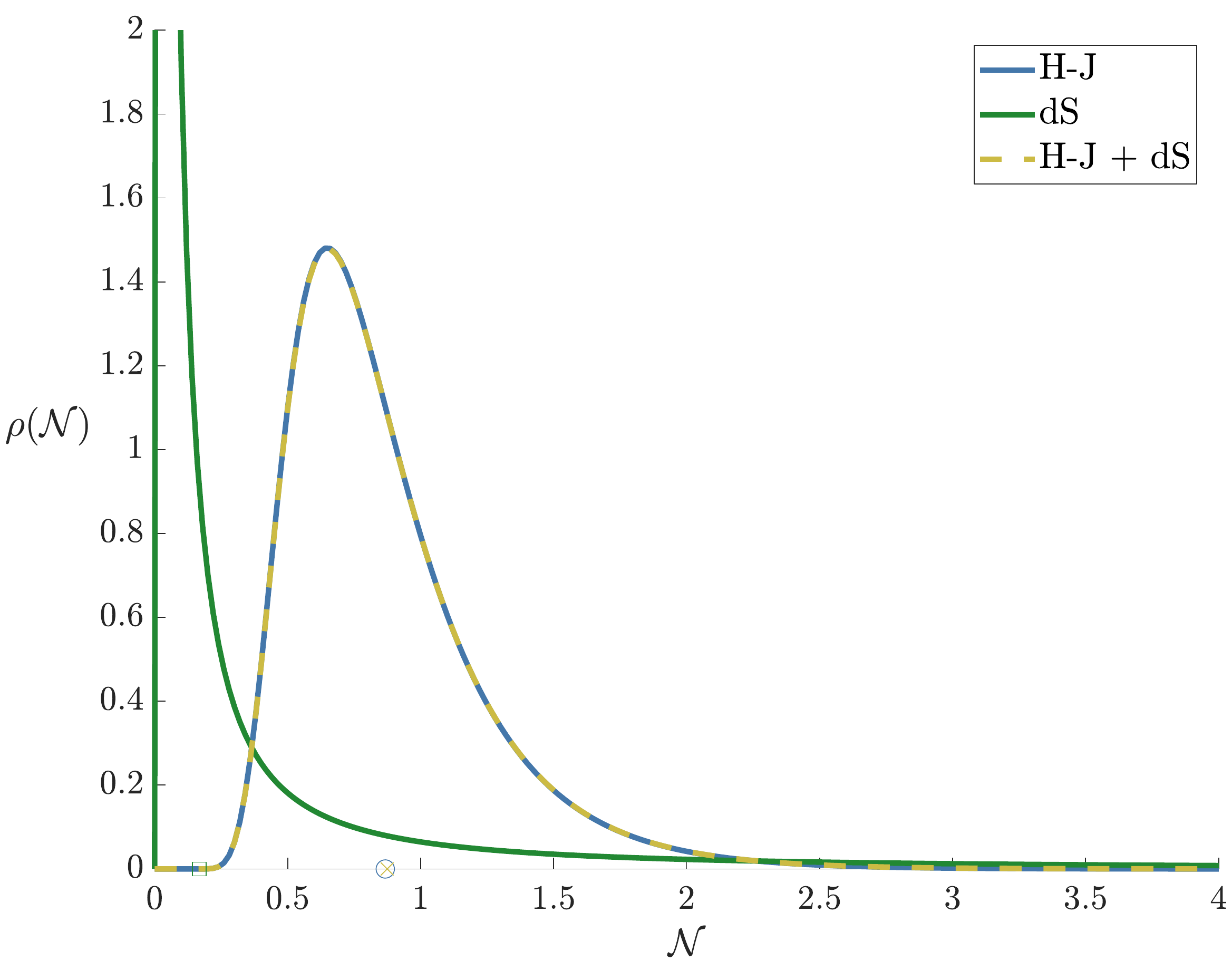}
\includegraphics[width=.48\textwidth]{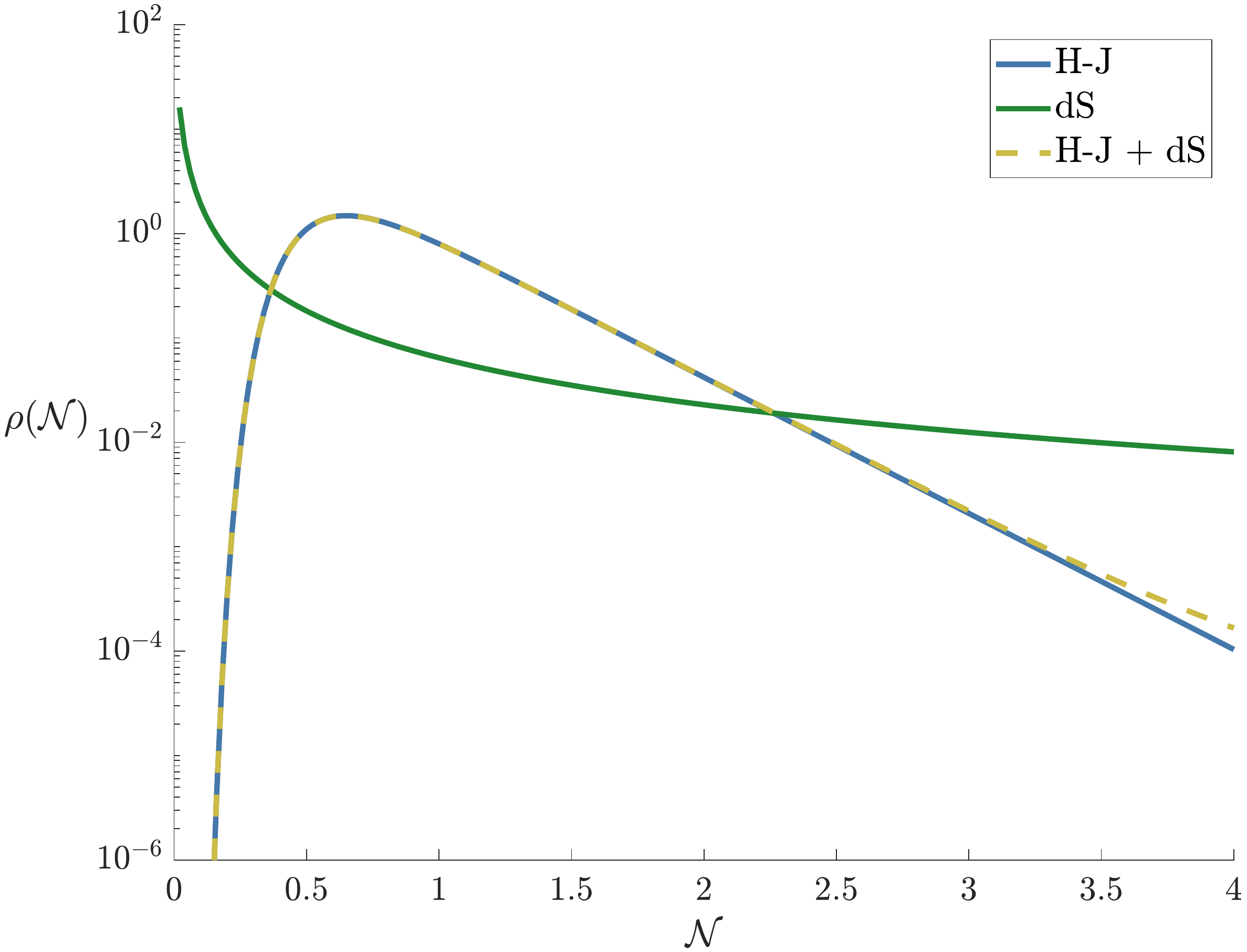}
\caption{\label{fig:rhoN scenarioB} The pdf for exit times, $\rho (\mathcal{N})$, in scenario B where $\chi = 5$, $\sigma = -10^{-2}$ plotted on a linear (left) and logarithmic scale (right). The green solid line corresponds to the pure free diffusion result (\ref{eq:PDF de Sitter}) and the solid blue to the H-J solution (\ref{eq:rho chie = 0}). The dashed line is the full H-J plus free diffusion phase (\ref{eq:rhoHJ+dS}). The circle, square and cross on the left plot correspond to the average number of e-folds, $\left\langle \mathcal{N}\right\rangle$, realised during the H-J phase, free diffusion only and H-J + free diffusion phase respectively - the circle and cross are essentially identical. Adding free diffusion to the H-J result only enhances the tail slightly as seen in the right plot. Note that both the average number of e-folds, the variance and the general shape of the PDF in the H-J computation differ significantly from those in the pure diffusion computation.}
\end{figure}

The analytical evaluation of the integral in (\ref{eq:rhoHJ+dS}) and the summation of the series is challenging, but a numerical evaluation is feasible. We show what the PDF looks like in Fig.~\ref{fig:rhoN scenarioB} for $\sigma=10^{-2}$, where the corresponding free diffusion PDF is also shown for comparison. As expected, for the very small values of $\sigma$ allowed, the total PDF resembles very closely the H-J one and increasing $\sigma$ only serves to slightly enhance the tail. From this PDF we can compute the total mass fraction of PBHs accounting for both the H-J and free diffusion phase. This is shown by the solid line for $\chi = 5$ on the right plot of Fig.~\ref{fig:Pattison muN and beta}. The other values of $\chi$ were not shown due to being indistinguishable graphically. The mass fraction is largely unchanged from its $\sigma = 0$ value for the plotted $\sigma$ range. Evaluating it for $|\sigma|>5 \times 10^{-2}$ is time-consuming and unnecessary  and the result would simply follow the dashed free diffusion line. The conclusion to be drawn however is clear: From the right plot of Fig.~\ref{fig:Pattison muN and beta} we see that $\beta$ in the $\sigma <0$ case is always above the allowed value - allowing for any period of free diffusion always overproduces black holes according to the H-J computation.

\section{\label{sec: Inflection}Abundance of PBHs from a local inflection point}
\begin{figure}[tbp]
\centering 
\includegraphics[width=.48\textwidth]{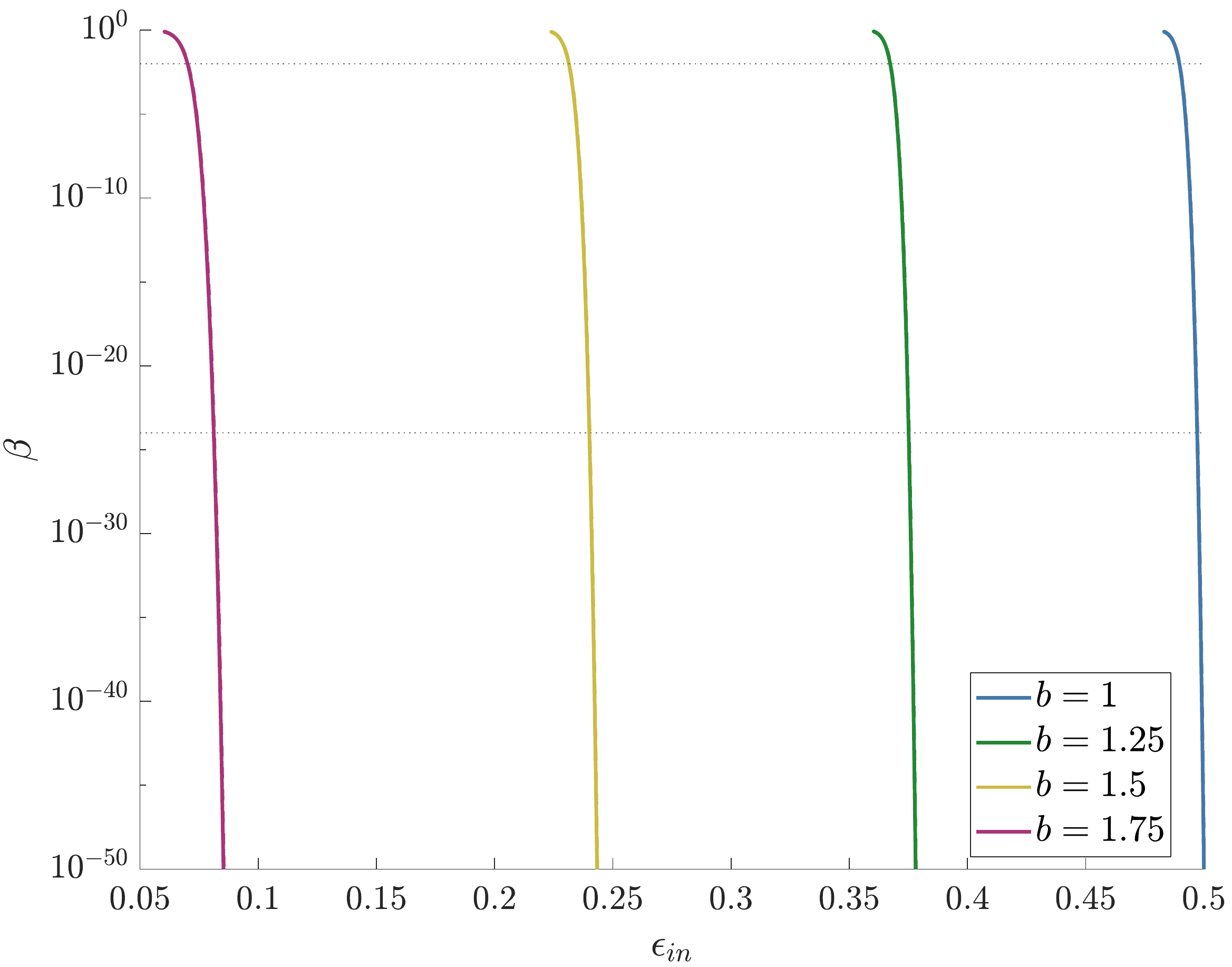}
\includegraphics[width=.48\textwidth]{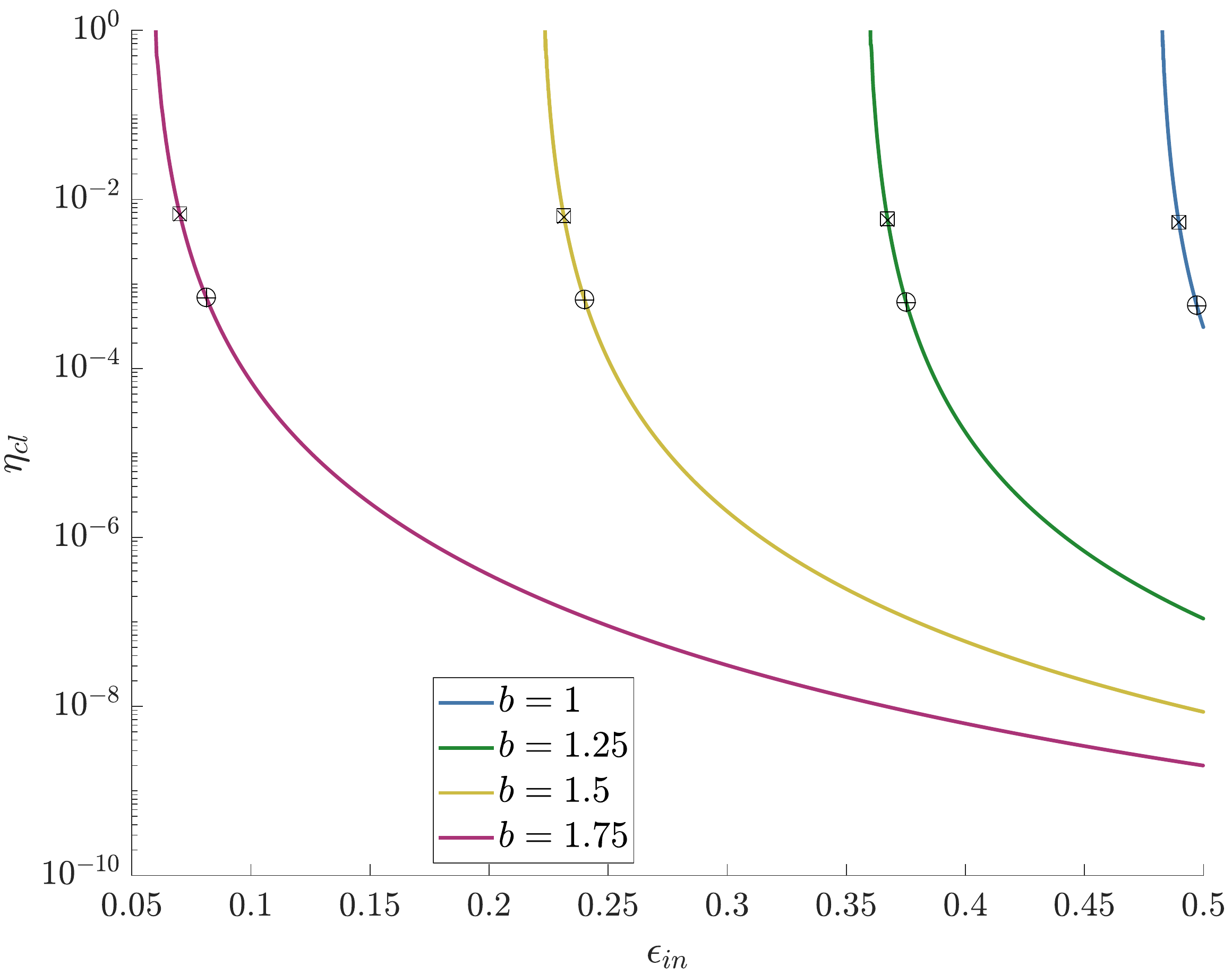}
\caption{\label{fig:inflec_powerandbeta.pdf} The left panel shows the mass fraction, $\beta$, as a function of the first Hubble slow-roll parameter as the inflaton enters the inflection point region for four different values of $b$. The solid and dashed lines represent the NLO (\ref{eq:beta NLO}) and NNLO (\ref{eq:beta NNLO}) approximations respectively and the dotted horizontal lines the weakest, $10^{-2}$, and strongest, $10^{-24}$ , constraints. The right panel shows the value of the classicality parameter $\eta_{cl}$ in the same parameter space. The boxes (circles) and crosses (pluses) correspond to the NLO and NNLO approximations respectively for $\beta$ being equal to the weakest (strongest) constraint $10^{-2}$ ($10^{-24}$). }
\end{figure}

We now consider a smoother entry into a USR regime which we will approximate locally as an inflection point. In contrast to other work on stochastic inflation and inflection points \cite{Ezquiaga2020, Pattison2021}, we will not define an ``effectively flat'' region around the inflection point and use our plateau results. Instead we will solve the H-J equation exactly and see if this gives us qualitatively different results than our conclusions for a plateau.  Concretely, we can imagine Taylor expanding around the inflection point, $\phi_{i}$, to obtain:
\begin{eqnarray}
V = V_0\left[ 1 + b\left( \phi-\phi_{i}\right)^3\right] \label{eq: inflection potential}
\end{eqnarray}
with $V_0$ corresponding to the height of the inflection point. While we cannot obtain an analytic solution for $H$ using this potential, it is straightforward enough to obtain numerically from (\ref{eq: H-J equation}) subject to an initial condition. We will parametrise a family of initial conditions in terms of the value of the first slow parameter, $\epsilon_1$, evaluated as the field enters our inflection point potential approximation (\ref{eq: inflection potential}). As we are looking to maximise stochastic effects we will choose the largest possible value of $V_0$ as allowed by the CMB -- see e.g. \cite{Akrami2020} -- which is determined by $v_0 = 10^{-10}$.

Pattison et.al \cite{Pattison2017} outline a program to compute the PDF -- and thus the mass fraction -- using characterstic function techniques. As discussed earlier, their formulae for slow-roll can be fully valid outside the slow-roll regime under the replacement $v \rightarrow \tilde{H}^2$. We will focus on the expansion of the characteristic function around the classical limit. At leading order every trajectory takes the same amount of time and there are no curvature perturbations. One must go to next-to-leading order (NLO) to obtain curvature perturbations which have a gaussian shape resulting in a mass fraction of:
\begin{eqnarray}
\beta_{\scalebox{0.5}{$\mathrm{NLO}$}} &=& \text{erfc}\left( \dfrac{\zeta_c}{2\tilde{H}\sqrt{\gamma_{1}^{\scalebox{0.5}{$\mathrm{NLO}$}}}} \right)\label{eq:beta NLO}\\
\gamma_{1}^{\scalebox{0.5}{$\mathrm{NLO}$}} &\equiv & \dfrac{1}{8\tilde{H}^2M_{Pl}^{4}}\int_{\phi_{end}}^{\phi_{in}}\mathrm{d}\phi~\dfrac{\tilde{H}^5}{\tilde{H}_{,\phi}^3} \label{eq:gamma1 NLO}
\end{eqnarray}
If we go to the next-to-next-to leading order (NNLO) we obtain non-gaussianties and a modified mass fraction:
\begin{eqnarray}
\beta_{\scalebox{0.5}{$\mathrm{NNLO}$}} &=& \text{erfc}\left( \dfrac{\zeta_c}{2\tilde{H}\sqrt{\gamma_{1}^{\scalebox{0.5}{$\mathrm{NNLO}$}}}} \right) + \dfrac{\gamma_{2}^{\scalebox{0.5}{$\mathrm{NNLO}$}}\left[\zeta_{c}^{2}-2\gamma_{1}^{\scalebox{0.5}{$\mathrm{NNLO}$}}\tilde{H}^2\right]}{4\tilde{H}\sqrt{\pi (\gamma_{1}^{\scalebox{0.5}{$\mathrm{NNLO}$}})^5}}\text{exp}\left(-\dfrac{\zeta_{c}^{2}}{4\tilde{H}^2\gamma_{1}^{\scalebox{0.5}{$\mathrm{NNLO}$}}}\right) \label{eq:beta NNLO} \\
\gamma_{1}^{\scalebox{0.5}{$\mathrm{NNLO}$}} &\equiv & \dfrac{1}{16\tilde{H}^2M_{Pl}^{4}}\int_{\phi_{end}}^{\phi_{in}}\mathrm{d}\phi \left[2\dfrac{\tilde{H}^5}{\tilde{H}_{,\phi}^3} + 7\dfrac{\tilde{H}^7}{\tilde{H}_{,\phi}^3} - 5\dfrac{\tilde{H}^8\tilde{H}_{,\phi\phi}}{\tilde{H}_{,\phi}^5}\right]\label{eq:gamma1 NNLO}\\
\gamma_{2}^{\scalebox{0.5}{$\mathrm{NNLO}$}} &\equiv & \dfrac{1}{16\tilde{H}^4M_{Pl}^{6}}\int_{\phi_{end}}^{\phi_{in}}\mathrm{d}\phi~\dfrac{\tilde{H}^9}{\tilde{H}_{,\phi}^5} \label{eq:gamma2 NNNO}
\end{eqnarray}
The mass fraction, $\beta$, in both the NLO and NNLO approximations is plotted on the left panel of Fig.~\ref{fig:inflec_powerandbeta.pdf}. We can see that $\beta_{\scalebox{0.5}{$\mathrm{NLO}$}}$ and $\beta_{\scalebox{0.5}{$\mathrm{NNLO}$}}$ are essentially indistinguishable in this regime.\\

As we are expanding around the classical limit we require that the classicality parameter (\ref{eq: classicality criterion}) $\eta_{cl}\ll 1$ for these formulae to be valid. In the right panel of Fig.~\ref{fig:inflec_powerandbeta.pdf} we plot the classicality parameter and show that for both the NLO and NNLO approximation the weakest bounds on $\beta$ are violated at around $\eta_{cl} \sim 10^{-2}$ and the strongest bounds are violated at around $\eta_{cl} \sim 10^{-3}$. In both cases $\eta_{cl} \ll 1$ which corroborates the conclusions of the previous section's plateau analysis. This demonstrates that \emph{PBHs will be generically overproduced before the inflaton can enter a quantum diffusion dominated regime} which corresponds to $\eta_{cl} \geq 1$. This is not to say that quantum diffusion effects aren't important and we expect them to play an important role in enhancing the tail of the distribution \cite{Ezquiaga2020}. While accounting for these effects is undoubtedly important to get a precise value of the mass fraction, we would only expect these effects to \textit{enhance} the abundance of PBHs from the NNLO computation and therefore our statement that the inflaton never enters a quantum diffusion dominated regime is still valid.

\section{\label{sec: Conclusion}Conclusion}

The main result of this work can be stated as follows: \\

\noindent\textit{The inflaton cannot enter a period of quantum diffusion dominated dynamics without first overproducing Primordial Black Holes}. \\

\noindent A semi-classical approximation seems to always be adequate for observationally allowed inflationary dynamics. We demonstrated this by considering the evolution of the inflaton entering both a finite width plateau as well as a local inflection point from a previous Slow-Roll (SR) phase. We have updated the results of \cite{Prokopec2019} to obtain the probability density function of e-fold exit times, $\mathcal{N}$, in the stochastic inflation formalism which are valid beyond slow roll and can describe for a finite width plateau, or more generally an ultra-slow-roll phase (USR), taking into account the velocity of the field as it slides in the USR region.

We showed that for a classical drift distance, $\Delta \phi_{cl}$, larger than the plateau width, $\Delta \phi_{pl}$, corresponding to the scenario where the classical inflaton momentum carries the field all the way through the plateau, the mass fraction of PBHs, $\beta$, closely resembles a step function. Unless $\Delta \phi_{cl} \approx  \Delta \phi_{pl}$, corresponding to extremely small values of $\sigma \lesssim 10^{-9}$, the mass fraction of PBHs produced is negligible. On the contrary if $\sigma$ is too small then PBHs will be overproduced violating observational and theoretical constraints. This very sharp transition between negligible production and over-production of PBHs occurs around $\sigma \sim \dfrac{\sqrt{\pi}}{4\chi}~e^{3\zeta_c} \simeq e^{3\zeta_c}\sqrt{\dfrac{\pi v_0}{8\epsilon_{in}}}$. This would indicate that PBHs will always be overproduced before $\sigma$ can reach $0$, meaning that the inflaton is observationally forbidden from getting stranded on the plateau and exploring it via pure quantum diffusion. Furthermore, PBHs are overproduced even before $\sigma = \sigma_{cl}$, corresponding to when quantum diffusion effects would be dominant even for an inflaton that is still classically drifting. 

We tested the robustness of these constraints by taking the explicit case of $\Delta \phi_{cl} = \Delta \phi_{pl}$, or $\sigma = 0$, where the mass fraction $\beta$ can be computed exactly. 
In this case, the mass function $\beta$ generically lies in the range $10^{-1}$ - $10^{-2}$ for all realistic values of the cutoff $\zeta_c$ (Unless $\chi$ is very small, corresponding to Plankcian inflationary energies, where $\beta \rightarrow 1$) that the mass fraction remained constant as $\chi$ was varied. We can therefore say that the case where the classical field momentum carries the field right to the edge of the plateau, $\Delta \phi_{cl} = \Delta \phi_{pl}$, will always overproduce PBHs. Consequently, the scenario where $\Delta \phi_{cl} < \Delta \phi_{pl}$, corresponding to a period of free diffusion, is also forbidden as this subsequent phase would only serve to \textit{enhance} the curvature perturbations. We verified this assumption by extending the free diffusion results of \cite{Prokopec2019} to a finite width plateau which confirmed that the $\Delta \phi_{cl} < \Delta \phi_{pl}$ case always overproduces PBHs. We therefore arrive at the conclusion stated above, namely that there can be no period of free diffusion during inflation without overproducing Primordial Black Holes.

When examining the more general setup of an inflection point we found that even in the gaussian case, PBHs are overproduced before the classicality parameter $\eta_{cl}$ is violated. This agrees with the plateau result and further suggests that the distortion of the classical relationship between field values and wavenumbers explored in \cite{Ando2020} is never realised and that a late period of quantum diffusion which spoils the CMB power spectrum is already ruled out by PBH considerations. We leave the verification and further exploration of this point to future work.

% Future work could involve computing the full PDF numerically of inflection point potentials. 

\acknowledgments

AW would like to thank Chris Pattison and Sam Young for helpful discussions about aspects of stochastic inflation and PBH formation respectively. AW is funded by the EPSRC under Project 2120421.

\appendix

\section{\label{sec: More Venin formulae}First Passage Time formulae in the Hamilton-Jacobi formalism}
Modifying the results of \cite{Vennin2015} to be valid outside of slow-roll using $v \rightarrow \tilde{H}^2$ we straightforwardly obtain the following results.\\
The average number of e-folds, $\left\langle\mathcal{N}\right\rangle$, is given by:
\begin{eqnarray}
\left\langle\mathcal{N}\right\rangle (\phi_{*}) = \int_{\phi_{end}}^{\phi_{*}}\dfrac{\mathrm{d}x}{M_{Pl}}\int_{x}^{\bar{\phi}}\dfrac{\mathrm{d}y}{M_{Pl}}\dfrac{1}{\tilde{H}^2(y)}\text{exp}\left[ \dfrac{1}{\tilde{H}^2(y)} - \dfrac{1}{\tilde{H}^2(x)} \right] \label{eq:mean N full}
\end{eqnarray}
The variation in the number of e-folds, $\delta \mathcal{N}^2 = \left\langle\mathcal{N}^2\right\rangle - \left\langle\mathcal{N}\right\rangle^2$ is given by:
\begin{eqnarray}
\delta \mathcal{N}(\phi_{*})^2 = \int_{\phi_{end}}^{\phi_{*}}\mathrm{d}x\int_{x}^{\bar{\phi}}\mathrm{d}y \left[\left\langle\mathcal{N}\right\rangle '(y)\right]^2\text{exp}\left[ \dfrac{1}{\tilde{H}^2(y)} - \dfrac{1}{\tilde{H}^2(x)} \right] \label{eq:Var N full}
\end{eqnarray}
The power spectrum:
\begin{eqnarray}
\mathcal{P}_{\zeta} = \dfrac{\mathrm{d}\delta\mathcal{N}^2}{\mathrm{d}\left\langle \mathcal{N}\right\rangle} \label{eq:Powerspecfull}
\end{eqnarray}
The local $f_{\mathrm{NL}}$ parameter:
\begin{eqnarray}
f_{\mathrm{NL}} = \dfrac{5}{72}\dfrac{\mathrm{d}\delta\mathcal{N}^3}{\mathrm{d}\left\langle \mathcal{N}\right\rangle}^2 \left(\dfrac{\mathrm{d}\delta\mathcal{N}^2}{\mathrm{d}\left\langle \mathcal{N}\right\rangle}\right)^{-2} \label{eq:fNL full}
\end{eqnarray}

\section{\label{sec: SR + USR}Convolving Slow-Roll and Ultra Slow-Roll}
The probability density for exit time given a SR phase followed by an USR phase is given by the convolution of the two PDFs:

\begin{eqnarray}
\rho_{SR+USR}(\mathcal{N}_2) = \int_{-\infty}^{\infty} d \mathcal{N}_1~\rho_{SR}(\mathcal{N}_1)\rho_{USR}(\mathcal{N}_2 -\mathcal{N}_1)
\end{eqnarray}
where the individual pdfs are given by:
\begin{eqnarray}
\rho_{SR}(\mathcal{N}_1) &=&  \dfrac{1}{\sigma_{SR} \sqrt{2\pi}}\text{exp}\left[ -\dfrac{1}{2}\dfrac{(\mathcal{N}_1 - \left\langle  \mathcal{N}_1\right\rangle)^2}{\sigma_{SR}^{2}}\right] \\
\rho_{USR}(\mathcal{N}_2 -\mathcal{N}_1) &=& \dfrac{6}{\sqrt{\pi}}\chi ~e^{-\chi^{2}\sigma^2}e^{-3\Delta \mathcal{N}_{12}}
\end{eqnarray}
where $\Delta \mathcal{N}_{12} \equiv \mathcal{N}_2 - \mathcal{N}_1$ and $\sigma_{SR} = \sqrt{v_0/\epsilon_{in}} \approx 1/\sqrt{2}\chi$. Performing the integration over $\mathcal{N}_1$ and realising that the lower bound of integration is restricted to $\left\langle \mathcal{N}_1 \right\rangle$ as $\rho_{USR}$ has zero weight below this:
\begin{eqnarray}
\rho_{tot}(\mathcal{N}_2) &=& \dfrac{3}{\sqrt{\pi}}\chi ~e^{-\chi^{2}\sigma^2} e^{-3(\mathcal{N}_2 - \left\langle \mathcal{N}_2\right\rangle )}e^{-3(\left\langle \mathcal{N}_2\right\rangle - \left\langle \mathcal{N}_1\right\rangle )} e^{\frac{9}{4}\chi^{-2}} \text{erfc}\left[ -\dfrac{3}{\sqrt{2}\chi} -\chi(\sqrt{2} -1)\left\langle \mathcal{N}_1\right\rangle \right] \nonumber \\
\\
&\approx &  \dfrac{6}{\sqrt{\pi}}\chi ~e^{-\chi^{2}\sigma^2} e^{-3(\zeta_c + \left\langle \mathcal{N}_2\right\rangle - \left\langle \mathcal{N}_1\right\rangle )} e^{\frac{9}{4}\chi^{-2}}
\end{eqnarray}
where in the second line we have approximated the complementary error function as 2 which is valid as the argument is generically large and negative. We can then use (\ref{eq:massfracdef}) to obtain the convolved PBH mass fraction:
\begin{eqnarray}
\beta_{SR+USR}(M) &=& \dfrac{4}{\sqrt{\pi}}\chi ~e^{-\chi^{2}\sigma^2}~e^{-3(\zeta_c +  \left\langle  N_2\right\rangle -\left\langle  N_1\right\rangle)} \times e^{\frac{9}{4}\chi^{-2}} \\
&=& \beta_{USR} \times e^{\frac{9}{4}\chi^{-2}} \label{eq:betaUSR+SR}
\end{eqnarray}
So we can see that a previous SR phase enhances the USR calculation by a factor of $e^{\frac{9}{4}\chi^{-2}}$. This factor is clearly negligible for $\chi \gg 1$ and is only relevant for $\chi < 1$ where it significantly enhances the mass fraction $\beta$.

\section{\label{sec: PDF for scenario B}Derivation of full PDF for scenario B}
In \cite{Prokopec2019} the PDF was computed for a free diffusion phase on an infinitely wide plateau with one absorbing boundary. Here we will adapt this computation to account for a finite plateau by including a reflecting boundary at one side in spirit with the computation in \cite{Pattison2017}.

We begin by recalling that the injected current $J(\alpha)$ into the free diffusion branch from the H-J phase is given by:
\begin{eqnarray}
J(\alpha) = \dfrac{6\pi\chi}{\left[ 2\pi\text{sinh}(3\Delta \alpha)\right]^{3/2}}~\text{exp}\left[\dfrac{3}{2}\Delta\alpha -\dfrac{1}{2}\left( \text{coth}(3\Delta \alpha ) -1\right)\chi^2\right] \label{eq:J current}
\end{eqnarray}
This will allow us to compute the probability distribution for $\phi$ on the free diffusion branch, $P_{v_0}$, through:
\begin{eqnarray}
P_{v_0}(\phi,\alpha) = \int_{\alpha_{in}}^{\alpha}\mathrm{d}u ~G(\phi-\phi_0, \alpha - u )J(u) \label{eq:Pv0}
\end{eqnarray}
where $G$ is the solution to the diffusive Green's function with exit boundary condition at $\phi_{e}$ and reflecting boundary condition at $\phi_{in}$:
\begin{eqnarray}
\partial_{\alpha}G = \tilde{H}_{0}^{2}\partial_{\phi\phi}G, \quad G(\phi = \phi_e , \alpha) = \partial_{\phi}G(\phi = \phi_{in} , \alpha) = 0,\quad G(\phi, \Delta\alpha \rightarrow 0) = \delta (\phi-\phi_0)\nonumber \\ \label{eq:greenseqn}
\end{eqnarray}
which has the solution:
\begin{eqnarray}
G(\phi,\Delta \alpha) &=& \dfrac{1}{\sqrt{2}\Delta\phi_{pl}}\sum_{n=0}^{\infty}\text{exp}\left[-\left(n + \dfrac{1}{2}\right)^2\dfrac{\tilde{H}_{0}^{2}}{\Delta\phi_{pl}^{2}}\pi^2\Delta\alpha \right]\nonumber \\
&& \times \left\lbrace \text{cos}\left[ \left(n + \dfrac{1}{2}\right)\dfrac{\pi}{\Delta\phi_{pl}}(\phi-\phi_0)\right] - \text{cos}\left[ \left(n + \dfrac{1}{2}\right)\dfrac{\pi}{\Delta\phi_{pl}}(\phi+\phi_0-2\phi_e)\right]\right\rbrace \nonumber \\\label{eq:G soln}
\end{eqnarray}
We can use this solution and equation (\ref{eq:Pv0}) to obtain the PDF for exit times, $\rho_{HJ+dS}(\mathcal{N})$, for a H-J phase followed by a free diffusion phase using the relation:
\begin{eqnarray}
\rho_{HJ+dS}(\mathcal{N}) = \tilde{H}_{0}^{2}\dfrac{\partial P_{v_0}(\phi,\mathcal{N})}{\partial \phi}\vert_{\phi_{e}} \label{eq:rhoHJ+dS defn}
\end{eqnarray}
yielding
\begin{eqnarray}
\rho_{HJ+dS}(\mathcal{N}) &=& \dfrac{2\sqrt{\pi}}{\chi(1-\sigma)^2}\int_{0}^{\mathcal{N}}\mathrm{d}u~\left[\text{sinh}(3u)\right]^{-3/2}~\text{exp}\left[\dfrac{3}{2}u -\dfrac{1}{2}\left( \text{coth}(3u ) -1\right)\chi^2\right] \nonumber \\
&&\times \sum_{n=0}^{\infty}\left(n+\dfrac{1}{2}\right)\text{sin}\left[\left((n+\dfrac{1}{2}\right)\dfrac{\pi\sigma}{\sigma-1}\right]\text{exp}\left[-\left(n + \dfrac{1}{2}\right)^2\dfrac{2\pi^2}{3\chi^2 (1-\sigma)^2} (\mathcal{N} -u) \right] \nonumber \\ \label{eq:append rhoHJ+dS}
\end{eqnarray}
Equation (\ref{eq:append rhoHJ+dS}) is the main result of this appendix.
%\paragraph{Note added.} This is also a good position for notes added after the paper has been written.


%\begin{thebibliography}{99}
\bibliographystyle{ieeetr85}
\bibliography{library}

% Please avoid comments such as "For a review'', "For some examples",
% "and references therein" or move them in the text. In general,
% please leave only references in the bibliography and move all
% accessory text in footnotes.

% Also, please have only one work for each \bibitem.

%\end{thebibliography}
\end{document}